\documentclass[aip,amsmath,amssymb,reprint]{revtex4-1}

\usepackage[utf8]{inputenc}
\usepackage{makecell}
\usepackage{array}
\usepackage{dcolumn}
\usepackage{physics}
\usepackage{lmodern}
\usepackage{graphicx}
\usepackage{amsmath}
\usepackage{amssymb}
\usepackage{gensymb}
\usepackage{float}
\usepackage{color}
\usepackage{xspace}
\usepackage[version=4]{mhchem}
\usepackage[T1]{fontenc}

\graphicspath{{figures/}}

\newcommand{\red}{}

\newcommand{\para}{\textit{para}-H$_2$\xspace}
\newcommand{\spara}{solid \textit{para}-H$_2$\xspace}

\newcommand{\hcp}{\textit{hcp}\xspace}

\newcommand{\pHthree}{(\textit{para}-H$_2$)$_3$\xspace}
\newcommand{\wvn}{cm$^{-1}$\xspace}

\newcommand{\eng}[2]{E^{#1}_{#2}}

\usepackage{silence}
\begin{document}


\title{Three-body potential energy surface for parahydrogen}

\author{Alexander Ibrahim}
\affiliation{Department of Physics and Astronomy, University of Waterloo, 200 University Avenue West, Waterloo, Ontario N2L 3G1, Canada}
\affiliation{Department of Chemistry, University of Waterloo, 200 University Avenue West, Waterloo, Ontario N2L 3G1, Canada}

\author{Pierre-Nicholas Roy}
\email{pnroy@uwaterloo.ca}
\affiliation{Department of Chemistry, University of Waterloo, 200 University Avenue West, Waterloo, Ontario N2L 3G1, Canada}


\begin{abstract}

We present a 3D isotropic \textit{ab initio} three-body \pHthree interaction potential energy surface (PES). The electronic structure calculations are carried out at the correlated coupled-cluster theory level, with single, double, and perturbative triple excitations. \red{The calculations use} an augmented correlation-consistent triple zeta basis set and a supplementary midbond function. We construct the PES using the Reproducing-Kernel Hilbert Space toolkit [J. Chem. Inf. Model. {\bf 57}, 1923 (2017)] with phenomenological and empirical adjustments to account for short-range and long-range behaviour. The \pHthree interaction energies deviate drastically from the Axilrod-Teller-Muto (ATM) potential at short intermolecular separations. We find that the configuration of three \para molecules at the corners of an equilateral triangle is responsible for the majority of the \pHthree interaction energy contribution in a hexagonal-close-packed lattice. In cases where two \para molecules are close to one another while the third is far away, the \pHthree interaction PES takes the form of a modified version of the ATM potential. We expect the combination of this PES together with a first principles \para--\para Adiabatic Hindered Rotor potential to outperform a widely-used effective pair potential for condensed many-body systems of \para.

\end{abstract}

\maketitle


\section{Introduction} \label{sec:intro}


Molecular hydrogen is the simplest of all molecular species, and is the target of a large body of fundamental research. In particular, many-body systems of one of its nuclear spin isomers, para-hydrogen (\para) display a vast array of interesting quantum behaviours. For example, there exist both experimental\cite{superfluidh2:00greb, superfluidh2:01greb:1, superfluidh2:01greb:2, superfluidh2:02greb, superfluidh2:03greb} and theoretical\cite{superfluidh2:91sind, pathinteg:02kwon, superfluidh2:05paes, superfluidh2:10li, superfluidh2:13zeng:1, superfluidh2:13zeng:1, superfluidh2:20boni} evidence of superfluid behaviour in pure and doped clusters of \para, or doped clusters of \para embedded within helium droplets. Solid \para has been established as a quantum solid.\cite{ph2solidexp:60gush, ph2solidtheo:66nosa, ph2solidexp:67bost, solidh2theo:83kran} In \spara, each molecule has a large zero-point motion about its lattice site (with a Lindemann ratio of about $0.2$),\cite{ph2solidexp:12fern, ph2solidtheo:17duss} owing to the low mass of, and weak intermolecular interactions between, \para molecules.\cite{ph2solidtheo:66nosa, ph2solidexp:80silv} These zero-point lattice vibrations are substantial enough to ``inflate'' the lattice, and give it many interesting quantum properties. In the zero-temperature, zero-pressure (ZTZP) limit, the lattice constant of \spara is $ R_0 = 3.79 $ \AA,\cite{ph2solidexp:80silv} considerably higher than what one would predict using a classical description. The combination of the spherical symmetry of its $ j = 0 $ rotational level, its weak intermolecular interactions, low chemical reactivity, and unsuspectingly large lattice constant, makes \spara a powerful host lattice for matrix isolation spectroscopy.\cite{matisosp:79wehr, matisosp:94barn, matisosp:96bond, matisosp:98momo, matisosp:05momo, matisosp:06yosh, matisosp:10baho} Molecules small enough to occupy single substitution sites in \spara, such as \ce{H2O},\cite{matisosp:04faja, matisosp:08faja} \ce{CH4},\cite{matisosp:00miki, matisosp:08miya} and \ce{CO},\cite{matisosp:09faja} are capable of nearly free rotation.

Similarly to how \para is the simplest molecular species, \spara is one of the simplest examples of a molecular solid. The intermolecular interactions are weak enough for the \para molecules within the crystal to retain many of their free molecule properties.\cite{ph2solidexp:80silv, solidh2theo:83kran} In fact, it is possible to study many of the solid's properties largely in terms of the free molecule properties. For example, except at very high pressures, the free vibrational and rotational quantum numbers $ \nu $ and $ j $ remain good quantum numbers in the solid phase, and thus the internal vibrations and rotations of the \para molecules remain ``free'' within the solid.\cite{solidh2theo:83kran} In light of this, it is apparent that we can make large strides in our understanding of \spara by studying the interactions between free molecular \para.


We take our first steps into studying many-body collections of \para by looking at the pair potential. In the initial attempts at creating an interaction potential, scientists would assume a physical form informed by theoretical and phenomenological knowledge, with several adjustable parameters. These parameters would then be determined from experimental results, such as molecular scattering cross section data,\cite{h2pes:72farr, h2pes:83buck} virial and viscosity coefficients,\cite{h2pes:54maso} gas transport properties,\cite{h2pes:48hirs} and solid state data.\cite{h2pes:78silv, h2pes:84norm} With advances in computing and electronic structure calculation research, many \ce{H2}--\ce{H2} interaction potentials were created based on \textit{ab initio} studies.\cite{h2pes:00diep, h2pes:08patk, h2pes:08hind, ph2cluster:14faru} One very commonly used and successful pair potential is the Silvera-Goldman (SG) potential,\cite{h2pes:78silv} a semiempirical potential that uses a combination of solid state data and self-consistent field calculations.


A more complete picture of bulk systems of \para molecules requires the inclusion of many-body potentials. Advances in experimental and computational technology have made it possible to get increasingly accurate quantitative descriptions of condensed systems, and so research into many-body interactions has intensified.\cite{manybody:92szcz, manybody:94elro} At long intermolecular separations, the triple-dipole interaction for spherical molecules is given by the Axilrod-Teller-Muto (ATM) potential\cite{threebody:43axil, threebody:43muto}
\begin{equation} \label{eq:atm_original}
    V_3(\{ R_{ij} \}, \{ \alpha_i \}) =
        C_9
        \left[
            \frac{1 + 3 \cos \alpha_1 \cos \alpha_2 \cos \alpha_3}{R_{12}^3 R_{13}^3 R_{23}^3}
        \right]
\end{equation}
\noindent
where $ C_9 $ is a coefficient that determines the interaction strength, and in the triangle formed by molecules $ 1 $, $ 2 $, and $ 3 $, $ R_{ij} $ is the side length connecting molecules $ i $ and $ j $, and $ \alpha_i $ is the internal angle at molecule $ i $. However, the \pHthree \red{three-body contribution to the interaction energy} deviates drastically from the ATM potential as the \para molecules move closer together. For most triangle configurations of the three molecules, including those most relevant to the \hcp lattice structure of \spara, the isotropic \pHthree \red{three-body contribution to the interaction energy} at short intermolecular spacings is attractive\cite{threebody:96wind, threebody:08hind}, whereas the ATM potential predicts a strong net repulsive \red{three-body contribution to the interaction energy}. Some works attempt to mitigate this short-range repulsion by applying damping functions to the ATM potential.\cite{damping:00sach, damping:10anat, damping:15huan} In addition, several \para pair potentials have been created that take many-body effects into account. For example, a potential by Moraldi\cite{h2pes:12mora} (and the modified version by Omiyinka and Boninsegni)\cite{h2pes:13omiy} softens the SG pair potential at short intermolecular distances, and is in excellent agreement with experiment. The SG potential has a two-body term that approximates the three-body interaction energy. Interestingly, this approximation term is repulsive.

\red{For brevity, from this point on we use the term ``\pHthree interaction energy'' to refer to the \pHthree three-body contribution to the total interaction energy, with no contribution from the pair interaction, unless otherwise specified.}

In 2010, Manzhos \textit{et al.} constructed an \textit{ab initio} 9D \pHthree potential energy surface (PES), with energies at the CCSD(T) level calculated using an aug-cc-pVTZ (AVTZ) atom-centred basis set.\cite{threebody:10manz} They sampled the positions and angular orientations of \ce{H2} molecules from a distribution, and fit the energies to a continuous functional form using a neural-network based algorithm. This PES gives the expected attractive behaviour at short intermolecular distances, and incorporates both intermolecular distance and angular orientation degrees of freedom. Garberoglio used a slightly modified version of this PES with much success to calculate the third virial coefficient of \para.\cite{virial:13garb} However, it was noted that the PES by Manzhos \textit{et al.} does not, without modifications, reproduce the correct behaviour at large intermolecular distances. Also, the black-box nature of the PES makes it more difficult to analyze; indeed, this was one of the factors that made it difficult to judge the accuracy of the third virial coefficient calculations.

Faruk \textit{et al.} produced a 1D first-principles \para--\para interaction potential,\cite{ph2cluster:14faru} by applying the Adiabatic Hindered Rotor method\cite{hinderedrotor:10li} to a 6D \ce{H2}--\ce{H2} PES published by Hinde.\cite{h2pes:08hind} We refer to this potential as the Faruk-Schmidt-Hinde (FSH) potential. The FSH potential was able to successfully reproduce experimentally observed vibrational energy shifts for parahydrogen clusters.\cite{ph2cluster:14faru, ph2cluster:15schm} In a recent paper, \red{path-integral Monte Carlo (PIMC)} simulations were performed using the FSH potential to generate an equation of state (EOS) of \spara.\cite{pathinteg:19ibra} The FSH potential was able to reasonably predict the ZTZP density and energy per particle, but greatly overestimated the pressure at higher densities compared to two other pair potentials, the SG and Buck\cite{h2pes:83buck} potentials. This is because the FSH potential does not account for the attractive three-body energies at short separations, and thus has a harder core than both the Buck and SG potentials.\cite{ph2cluster:15schm} The inclusion of a three-body \pHthree interaction potential alongside the FSH potential would thus decrease the pressure of \spara at higher densities, likely bringing it closer to experiment. \red{Janssen and Avoird\cite{ph2solidtheo:90jans} studied the dynamics and phase transitions of solid \para and \textit{ortho}-D$_2$ using an \textit{ab initio} pair potential, and concluded that three-body interactions are needed to correctly predict the solid's behaviour at high pressures.}

We construct a 3D \textit{ab initio} isotropic PES for the \pHthree trimer system. Our calculations are carried out at the CCSD(T) level, using the AVTZ atom-centred basis set, with an additional $ (3s3p2d) $ bond function.\cite{elecstr:92tao} We apply the rigid rotor approximation, and spherically average over the angular orientations of the \ce{H2} molecules. The energies are fit to a smooth PES using the Reproducing-Kernel Hilbert Space (RKHS) toolkit by Unke \textit{et al.}\cite{rkhs:17unke} The RKHS method is a machine learning method, primarily used to construct multidimensional PESs from \textit{ab initio} data,\cite{rkhs:96ho, rkhs:03ho} and has successfully been used to create PESs for several other molecular systems.\cite{rkhsex:96ho, rkhsex:02ho, rkhsex:14carl, rkhsex:16unke} This PES is capable of calculating accurate energies for all \pHthree systems that form triangle with all side lengths greater than $ 2.2 $ \AA, and offers suitable extrapolations to calculate energies for smaller triangles. At large intermolecular separations, the PES smoothly transitions to the ATM potential. For the case where two molecules are less than $ 3.2 $ \AA \, apart and the third is a large distance away, the PES smoothly transitions to a ``modified'' adaption of the ATM potential.

The remainder of this paper is structured as follows. In Sec.~\ref{sec:construct}, we describe the coordinate system used to represent our \pHthree interaction potential, the \textit{ab initio} calculations used to find the energies, and the methods used to construct the PES. In Sec.~\ref{sec:discuss}, we assess the errors in the PES due to basis set size, spherical averaging, and sparsity in the \textit{ab initio} input data mesh. We also look at the implications of using the presented PES in a study of \spara, and compare the presented PES against predictions made by the SG potential. In Sec.~\ref{sec:conclusion}, we present our conclusions and plans for future work.


\section{Creating the Potential Energy Surface} \label{sec:construct}
\subsection{Coordinate System of the Potential Energy Surface} \label{sec:construct:funcform}

Label the molecules with numbers $1$, $2$, and $3$. Let $ R_{ij} $ be the distance between the centres of mass of molecules $ i $ and $ j $. The angular orientation of molecule $ i $ is given by the space-fixed polar and azimuthal angles $ (\theta_i, \phi_i) $. We apply a rigid rotor approximation, fixing the bond length of each molecule to the vibrationally averaged \ce{H2} ground state ($\nu = 0$, $j = 0$) interatomic distance of $ 1.449 \ a_0 $. The rigid rotor approximation significantly reduces both the number of electronic structure calculations required to create the PES and the cost of calculating the energy in a simulation. Also, because the three-body energy is a correction to a much larger two-body (\ce{H2})$_2$ energy, the overall error due to the lack of bond length variations is expected to be small.

At fixed intermolecular distances $ \{ R_{ij} \} $, we apply the $6$-point Lebedev quadrature to spherically average over the angular orientations of the three \ce{H2} molecules.\cite{lebedev:76lebe, lebedev:03wang} More specifically, we align each \ce{H2} molecule along either the space-fixed $x$-, $y$-, or $z$-axis. Done independently for all three molecules, there are $27$ different combinations of space-fixed angles, and the \pHthree interaction energy in our PES is the average of the interaction energies for these $27$ different angular combinations. This averaging projects out the anisotropic components of the interaction energy, leaving only the isotropic term. An analysis of the error associated with using the $6$-point Lebedev quadrature as opposed to a higher-order Lebedev quadrature is discussed in Sec.~\ref{sec:discuss:lebedev}. Having removed both the bond length and angle degrees of freedom, each \ce{H2} molecule is treated as a point particle. The potential is now a function of the intermolecular distances, or equivalently, of the triangle formed by placing a \para molecule at each of its corners. From this point on, for brevity, when we refer to the ``distance between two molecules'', we mean the distance between their centres of mass, and when we refer to the ``interaction energy of a triangle configuration'', we refer to the interaction energy of the triangle where the centre of mass of a \para molecule lies at each corner.

\begin{figure}[h]
	\includegraphics[width=1.0\linewidth]{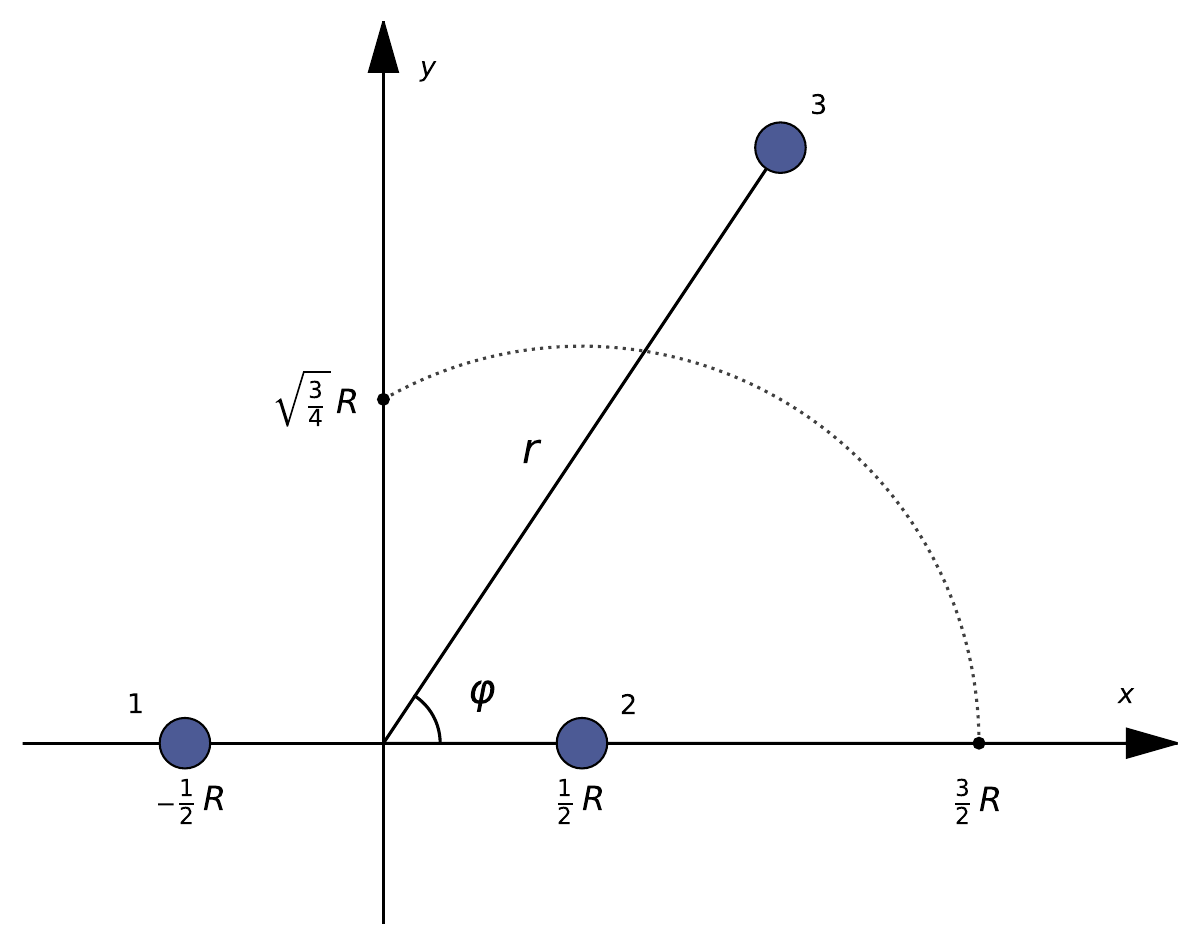}
    \caption{The centre-of-mass positions of the three \para molecules on the $ (x, y) $-plane. After removing the bond length and angular degrees of freedom, each \para molecule can be represented as a point. Molecule $3$ must lie outside of the region enclosed by the dotted line to satisfy the condition that $ R_{12} \leq R_{23} \leq R_{13} $.}
	\label{fig:paper1_fig1}
\end{figure}

In the present work, we use a ``rescaled'' Jacobi coordinate system to describe the relative centres of mass of the three \para molecules. First, label the molecules such that the pairwise distances satisfy $ R_{12} \leq R_{23} \leq R_{13} $. Define $ R = R_{12} $, and place molecules $1$ and $2$ on the $x$-axis, at $ (-R/2, 0) $ and $ (R/2, 0) $, respectively. Place molecule $3$ in the positive quadrant of the $ (x, y) $-plane. Draw a line connecting the origin to the position of molecule $3$. Define $ r $ as the length of this line, and define $ \varphi \in [ 0, \pi/2 ] $ as the angle between the positive $x$-axis and this line. This three-body coordinate system of \para molecules is depicted in Fig.~(\ref{fig:paper1_fig1}). This coordinate system is analagous to representing the closest two molecules as a dimer of bond length $ R $, while the third molecule is a monomer whose location in space is described relative to the centre of mass and angular orientation of the dimer. 

The variables $ R $, $ r $, and $\varphi $ can describe all triangles formed by the three \para molecules. However, due to the requirement that $ R \leq R_{23} \leq R_{13} $, the allowed values of $ r $ are coupled to both $ R $ and $ \varphi $. The minimum value of $ r $ is derived by setting $ R_{23} = R $, and is given by
\begin{equation} \label{eq:minimum_value_of_r}
    r_{\rm min} = \frac{R}{2} \left[ \cos\varphi + \sqrt{3 + \cos^2\varphi } \right] .
\end{equation}
\noindent
We decouple the coordinates $ R $ and $ \varphi $ from $ r $ by using the dimensionless variable $ s~=~r / r_{\rm min} $. The coordinate system used in the remainder of this paper is described by the independent variables $ (R, s, \varphi) $, where $ R \ge 0 $, $ s \ge 1 $, and $ \varphi \in [0, \pi/2] $. This ``rescaled'' Jacobi coordinate system is a convenient way of representing a PES for a three-body system of identical particles. Having three independent, non-repetitive coordinates makes it easy to set up a regular 3D grid. In addition, there is a simple, and unique, geometric interpretation associated with each coordinate from its interpretation as a dimer-monomer system. For example, the coordinate $ R $ represents both the bond length of the dimer and the overall size of the triangle, and changing it while keeping $ s $ and $ \varphi $ constant is akin to ``rescaling'' both the length of the dimer and the distance of the monomer from the dimer's centre by the same factor. Changing $ s $ while keeping $ R $ and $ \varphi $ constant is akin to moving the monomer towards or away from the centre of the dimer. The angle $ \varphi $ represents the orientation of the monomer's location relative to the long axis of the dimer.

Triangles of specific forms have convenient representations in the rescaled Jacobi coordinate system. As examples, the coordinates $ (R, 1, \pi/2) $, $ (R, 1, \tan^{-1}(1/2)) $, and $ (R, 1, 0) $ represent the equilateral triangle of side length $ R $, the right-angled triangle of leg length $ R $, and the case where the molecules lie on a straight line with a spacing of $ R $. The coordinates $ (R_0, s, \pi/2) $ represent the set of isosceles triangles of base length $ R_0 $ and height $ r \ge \sqrt{3/4} R_0 $. All triangles with the same value of $ s $ and $ \varphi $ are similar triangles.


\subsection{Electronic Structure Calculations} \label{sec:construct:abinitio}

We perform electronic structure calculations for the (\ce{H2})$_3$ interaction energy using the MRCC (version 2019) program.\cite{elecstr:13roli} The calculations are done with the coupled-cluster method with single, double, and perturbative triple excitations, abbreviated as CCSD(T).\cite{elecstr:89ragh} We use an AVTZ atom-centred basis set for each of the six hydrogen atoms, and supplement this with an additional set of $ (3s3p2d) $ midbond functions\cite{elecstr:92tao} positioned at the center of mass of the system, following the example by Hinde.\cite{h2pes:08hind} The \pHthree calculations by Wind \textit{et al.} and Manzhos \textit{et al.} did not make use of midbond functions.\cite{threebody:96wind, threebody:10manz} It has been shown that the AVTZ basis set provides similar energies to the aug-cc-pVQZ (AVQZ) basis set for the (\ce{H2})$_3$ trimer.\cite{threebody:96wind, threebody:08hind}

For a fixed coordinate $ (R, s, \varphi) $, let $ n $ label each of the $ 27 $ angular configurations of the $6$-point Lebedev quadrature. The counterpoise-corrected\cite{elecstr:70boys} (\ce{H2})$_3$ trimer interaction energy of the $ n^{\rm th} $ angular orientation (with $ (R, s, \varphi) $ omitted on the RHS for brevity) is calculated using
\begin{eqnarray} \label{eq:three_body_energy}
	\eng{n}{}(R, s, \varphi) = \eng{n}{123} - (\eng{n}{12[3]} + \eng{n}{13[2]} + \eng{n}{23[1]}) \nonumber \\ + (\eng{n}{1[23]} + \eng{n}{2[13]} + \eng{n}{3[12]}) \ .
\end{eqnarray}
\noindent
In the above equation, $ \eng{n}{123} $ is the total CCSD(T) energy of the trimer in the $ n^{\rm th} $ angular configuration. We define $ \eng{n}{ij[k]} $ to be the CCSD(T) energy of the $ i^{\rm th} $ and $ j^{\rm th} $ molecules in the complete one-electron basis set of the trimer. In other words, $ \eng{n}{ij[k]} $ is the CCSD(T) energy of the (\ce{H2})$_3$ trimer where the charge of the $ k^{\rm th} $ molecule is removed but \red{its} associated basis functions remain. Similarly, $ \eng{n}{i[jk]} $ is the CCSD(T) energy of the $ i^{\rm th} $ molecule in the complete one-electron basis set of the trimer. The spherically-averaged \textit{ab initio} \pHthree interaction energy $ V_{\rm CC} $ is calculated using
\begin{equation} \label{eq:lebedev_average}
    V_{\rm CC}(R, s, \varphi) = \frac{1}{27} \sum_{n = 1}^{27} \eng{n}{} (R, s, \varphi) \ .
\end{equation}

To create the input data mesh for the PES, we calculate $ V_{\rm CC} $ for $21$ values of $ R /$\AA $ \ \in [2.20, 6.25] $, $17$ values of $ s \in [1.0, 3.85] $, and $19$ values of $ \varphi \in [0, \pi/2] $. The spacing of $ \varphi $ values is uniform, while the spacings for $ R $ and $ s $ are non-uniform, with a denser selection of points at smaller values and a sparser selection at larger values. The error analysis associated with the spacing of points in the input data mesh is done in Sec~\ref{sec:discuss:sparsity}.

The precision of the CCSD(T) energies calculated by the MRCC program are determined by several convergence criteria. The calculations for these energies were performed with all default settings (for the 2019 version), except for \texttt{cctol}, which is set to \texttt{9}. We perform sample calculations using the AVTZ basis set using \texttt{cctol = 10}, for two triangular configurations. First, we do so for the equilateral triangle configuration $ (R, 1, \pi/2) $, ranging from $ R /$\AA $ = 2.20 $ to $ 6.25 $. All the sample interaction energies change by less than $ 0.0006 $~\wvn. At shorter intermolecular spacings, this is much smaller than the difference between \pHthree energies calculated using the AVTZ and AVQZ atom-centred basis sets. Next, we do so for the $ (2.2 \, \mathrm{\AA}, s, 0) $ configuration, where two molecules are a distance $ 2.2 $ \AA \, apart on the $x$-axis, and the third molecule moves away from them along the $x$-axis. We see the same behaviour. At intermolecular separations great enough where average \pHthree interaction energy becomes comparable to the error, the PES will have already been adjusted to an empirical form, as we shall see in Secs.~\ref{sec:construct:makepes:large_r} and \ref{sec:construct:makepes:large_s}.


\subsection{Constructing the Potential Energy Surface} \label{sec:construct:makepes}

We construct a smooth PES using a combination of several methods. First, we use the RKHS toolkit provided by Unke \textit{et al.} to construct the ``initial'' PES using the CCSD(T) energies.\cite{rkhs:17unke} The RKHS method has several convenient properties. For example, the resultant PES is smooth everywhere, and reproduces all the input CCSD(T) energies exactly. While the RKHS method is used for interpolating interaction energies within the $ (R, s, \varphi) $ input data mesh, we use a number of other strategies to extrapolate the PES outside of the grid.


\subsubsection{Large R Extrapolation} \label{sec:construct:makepes:large_r}

For cases where all three molecules are far apart ($ R_{ij} > 5.5 $ \AA), the \textit{ab initio} \pHthree interaction energies approach the ATM potential. Expressed using $ (R, s, \varphi) $ coordinates, the ATM potential is
\begin{equation} \label{eq:atm_potential_rsvarphi}
    V_{\rm ATM}(R, s, \varphi) = C_9 \frac{64}{R^9} \frac{1 + f(s, \varphi)}{\left[ T(s, \varphi) \right]^{3/2}}
\end{equation}
\noindent
where, using
\begin{equation} \label{eq:W_atm}
    W(s, \varphi) = s^2 \left[ \cos\varphi + \sqrt{3 + \cos^2\varphi} \right]^2 = 4\frac{r^2}{R^2} ,
\end{equation}
\noindent
we have
\begin{equation} \label{eq:T_atm}
    T(s, \varphi) = 1 - 2 W(s, \varphi) \cos2\varphi + W^2(s, \varphi)
\end{equation}
\noindent
and
\begin{equation} \label{eq:f_atm}
    f(s, \varphi) = - 3 \frac{(1 - W(s, \varphi) \cos^2\varphi)(1 - W(s, \varphi))}{T(s, \varphi)} .
\end{equation}
\noindent
We use the coefficient $ C_9 = 34336.220 $ \wvn \AA$^{9}$. \cite{threebody:08hind, moleculeh2:05hind}

\begin{figure}[h]
	\includegraphics[width=1.0\linewidth]{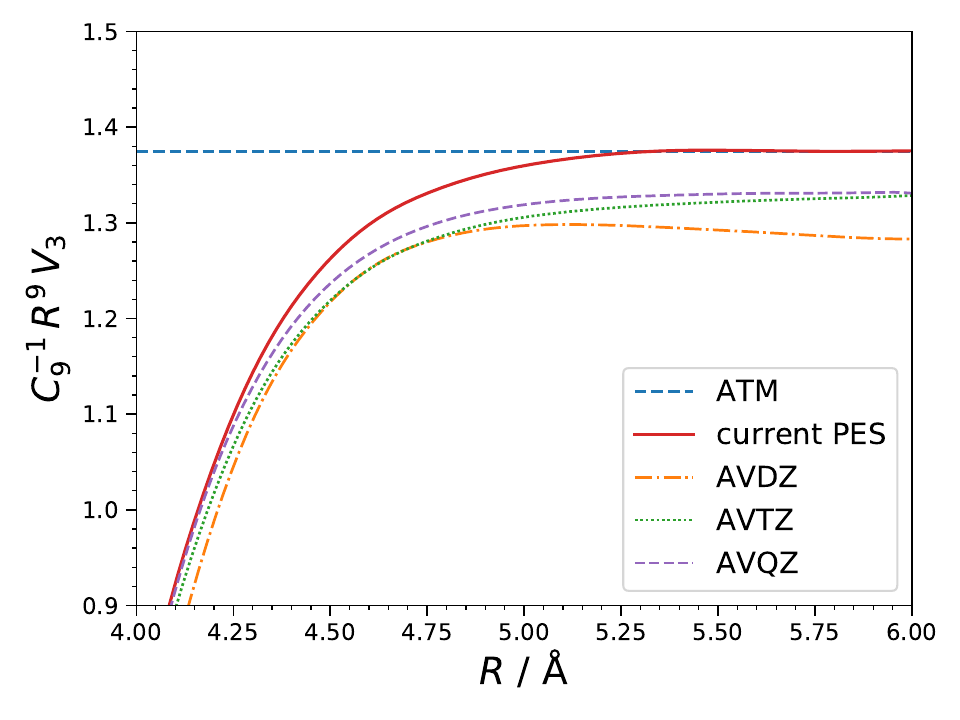}
    \caption{The \pHthree interaction energy for three \para molecules at the corners of an equilateral triangle of side length $ R $, multiplied by $ C_9^{-1} \, R^9 $. Shown are the ATM potential (horizontal, blue, dashed), and the RKHS-constructed \textit{ab initio} CCSD(T) energies calculated using the AVDZ (orange, dash-dotted), AVTZ (green, dotted), and AVQZ (purple, dashed) basis sets. Also shown is the current PES, constructed by applying the adjustments of Eq.~(\protect\ref{eq:extrap_large_R_taper}) to the AVTZ \textit{ab initio} energies, with $ R_{\rm A} = 5.35 $ \AA \, and $ R_{\rm C} = 3.6 $ \AA.}
	\label{fig:paper1_fig2}
\end{figure}

The \textit{ab initio} \pHthree interaction energies calculated using the AVDZ, AVTZ, and AVQZ basis sets do not precisely converge to the ATM potential at long distances. We ensure a smooth transition from the \textit{ab initio} AVTZ CCSD(T) energies to the ATM potential using the following procedure. First, we fix a pair of values $ s = \overline{s} $ and $ \varphi = \overline{\varphi} $ from the input data mesh. Then we plot both the ATM potential and the set of \textit{ab initio} \pHthree interaction energies $ \{ V_{\rm CC}(R; \overline{s}, \overline{\varphi}) \} $ against $ R $, and rescale all the energies by $ C_9^{-1} R^9 $ (see Fig.~(\ref{fig:paper1_fig2})). With this rescaling, the ATM potential appears as a horizontal line. Next we pick a value of $ R $ from the input data mesh, call it $ R_{\rm A} $, beyond which the rescaled \textit{ab initio} energies appear to flatten, indicating that the PES has begun its $ R^{-9} $ decay. We intend to adjust the PES such that $ V_3(R_{\rm A}; \overline{s}, \overline{\varphi}) = V_{\rm ATM}(R_{\rm A}; \overline{s}, \overline{\varphi}) $. We also pick a value $ R = R_{\rm C} $, not too far below $ R_{\rm A} $, at which the ATM and \textit{ab initio} energies have very clearly diverged when looking at the rescaled plot. The resultant PES energies $ \{ V_3(R; \overline{s}, \overline{\varphi}) \} $ are calculated (omitting $ \overline{s} $ and $ \overline{\varphi} $ on the RHS) using
\begin{equation} \label{eq:extrap_large_R_taper}
    V_3(R; \overline{s}, \overline{\varphi}) =
        \begin{cases}
            V_{\rm ATM}(R)                                       & R \ge R_{\rm A}, \\
            V_{\rm CC}(R) \left[ 1 + \eta(R, R_{\rm A}) \right]  & R_{\rm C} \le R < R_{\rm A}, \\
            V_{\rm CC}(R)                                        & R < R_{\rm C}
        \end{cases}
\end{equation}
\noindent
where
\begin{equation}
    \eta(R, R_{\rm A}) = \left[ \frac{V_{\rm ATM}(R_{\rm A}) - V_{\rm CC}(R_{\rm A})}{V_{\rm CC}(R_{\rm A})} \right] \, e^{-\lambda(R - R_{\rm A})^2 } ,
\end{equation}
\noindent
and $ \lambda = \lambda(\overline{s}, \overline{\varphi}) $, often set between $2$ and $5$, determines the rate of decay. In other words, we use the relative difference between the CCSD(T) and ATM energies at the chosen distance $ R_{\rm A} $ to smoothly taper the PES towards the ATM energies, for $ R_{\rm C} < R < R_{\rm A} $. To ensure the PES remains smooth, we apply these changes to the data level and reapply the RKHS method. As an example of the aforementioned procedure, in Fig.~(\ref{fig:paper1_fig2}) for the equilateral triangle configuration, we show the AVDZ-, AVTZ-, and AVQZ-level \textit{ab initio} \pHthree energy curves, the ATM potential, and the final adjusted PES. We see that the unadjusted \textit{ab initio} energies do not properly converge to the ATM potential at long distances; each underestimates the effective value of $ C_9 $ by about $ 3\% $. The current PES, constructed by applying Eq.~(\ref{eq:extrap_large_R_taper}) to the AVTZ curve, smoothly transitions to the ATM potential.


\subsubsection{Large s Extrapolation For R < 3.2 \AA} \label{sec:construct:makepes:large_s}

For cases where two molecules are close to one another ($ R < 3.2 $ \AA), but the third molecule is farther away (large $s$), the electronic structure calculations do not approach the ATM potential. Instead, \textit{ab initio} energies in the ``small $ R $, large $ s $'' regime converge to a modified ATM potential (the ``MATM potential''), which takes the form
\begin{equation} \label{eq:modatm_potential_rsvarphi}
    V_{\rm MATM}(R, s, \varphi) = C_9 \frac{64}{R^9} \frac{a(R, \varphi) + b(R, \varphi) \, f(s, \varphi)}{\left[ T(s, \varphi) \right]^{3/2}} \ .
\end{equation}
\noindent
Compared to Eq.~(\ref{eq:atm_potential_rsvarphi}), the only difference is the appearance of coefficients $ a(R, \varphi) $ and $ b(R, \varphi) $ in the numerator.

\begin{figure}[h]
	\includegraphics[width=1.0\linewidth]{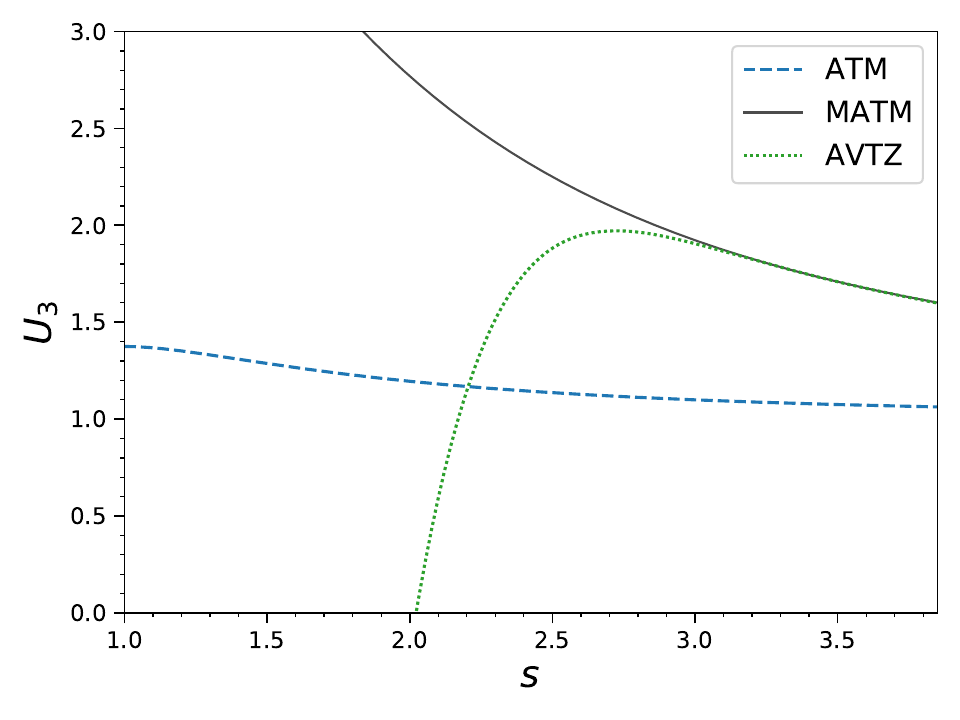}
    \caption{\red{The numerator term $ U_3(s; R, \varphi) = a + b \, f(s, \varphi) $, for $ (R, \varphi) = (2.2 \, \mathrm{\AA}, \pi/2) $.} We show the curves for the ATM potential (blue, dashed), the RKHS-constructed AVTZ-level \textit{ab initio} energies (green, dotted), and the MATM potential constructed to fit these \textit{ab initio} energies. In this case, $ a = 1.06 $ and $ b = 8.64 $.}
	\label{fig:paper1_fig3}
\end{figure}

We calculate $ a(R, \varphi) $ and $ b(R, \varphi) $ as follows. Define $ U_3(s; R, \varphi) = a + b \, f(s, \varphi) $; for the \textit{ab initio} energies, we find $ U_3(s; R, \varphi) $ by multiplying the energies by $ 64^{-1} C_9^{-1} R^9 T^{3/2}(s, \varphi) $. First, we fix the values of $ R = \overline{R} $ and $ \varphi = \overline{\varphi} $ from the input data mesh. Then, we pick some $ s^{\prime} $ for which $ U_3(s; \overline{R}, \overline{\varphi}) $ has transitioned to its long-range MATM behaviour for $ s \ge s^{\prime} $, and calculate $ a(\overline{R}, \overline{\varphi}) $ and $ b(\overline{R}, \overline{\varphi}) $ using
\begin{equation} \label{eq:a_and_b}
    b = \eval{ \frac{\partial U_3 / \partial s}{\partial f / \partial s} }_{s^{\prime}} \ \ \ \ \mathrm{and} \ \ \ \ a = U_3(s^{\prime}) - b \, f(s^{\prime}) \ .
\end{equation}
\noindent
For $ R > 3.2 $ \AA, the coefficients $ a(R, \varphi) $ and $ b(R, \varphi) $ both converge to unity and take the form of the nominal ATM potential. To give the constructed PES the correct long-range behaviour along the $s$-coordinate, we construct a new input mesh using
\begin{equation}
    V_3(s; \overline{R}, \overline{\varphi}) =
        \begin{cases}
            V_{\rm CC}(s; \overline{R}, \overline{\varphi})       & s \le s^{\prime} \\
            a(\overline{R}, \overline{\varphi}) + b(\overline{R}, \overline{\varphi}) \, f(s; \overline{\varphi})     & s >   s^{\prime}
        \end{cases}
\end{equation}
\noindent
where $ s^{\prime} $ is the value of $ s $ used to define the MATM potential in Eq.~(\ref{eq:a_and_b}). To ensure the transition occurs smoothly, we reapply the RKHS method on this new input mesh. In Fig.~(\ref{fig:paper1_fig3}), we plot $ U_3(s) $ for $ (R, \varphi) = (2.2 \mathrm{\ \AA}, \pi/2) $. We see that the \textit{ab initio} \pHthree interaction energies do not converge to the ATM potential when two \para molecules are close together. Instead, in this regime we fit the MATM potential to the long-range behaviour of the \textit{ab initio} energies along the $ s $-coordinate.

An unfortunate effect of using Eq.~(\ref{eq:a_and_b}) to calculate $ a(R, \varphi) $ and $ b(R, \varphi) $ is that both coefficients blow up as $ \varphi $ approaches zero. To see why, note that the partial derivative of Eq.~(\ref{eq:f_atm}),
\begin{equation} \label{eq:derivative_of_f}
    \pdv{f}{s} = \sin^2\varphi \, \frac{6W}{s} \left[ \frac{3 + 2W - (1 + 4 \cos^2\varphi)W^2}{T^2} \right] ,
\end{equation}
\noindent
is present in the denominator of Eq.~(\ref{eq:a_and_b}), and that, near $ \varphi = 0 $, $ b $ blows up as $ \sin^{-2}\varphi $. To work around this singularity, we perform the MATM fit of \textit{ab initio} energies along $ (\varphi = 0) $ using $ f(s, \varphi_{\rm cf}) $, with the cutoff $ \varphi_{\rm cf} = 10^{-5} $. This approximation has no effect whatsoever on the unextrapolated \textit{ab initio} energies. Performing the fit using $ \varphi_{\rm cf} = 10^{-3} $ or $ \varphi_{\rm cf} = 10^{-4} $ shows that this has little effect on the extrapolated energies, because as long as $ \varphi_{\rm cf} \neq 0 $, the coefficients $ a $ and $ b $ simply adjust themselves to maintain the MATM fit.

When $ R < 3.2 $ \AA, we can use Eq.~(\ref{eq:modatm_potential_rsvarphi}) to calculate \pHthree interaction energies when $ s > 3.85 $. For most values of $ \varphi $, we calculate coefficients $ a $ and $ b $  using
\begin{equation} \label{eq:realtime_calculate_a}
    a = \frac{U_3(s_0)f(s_1) - U_3(s_1) f(s_0)}{f(s_1) - f(s_0)}
\end{equation}
\noindent
and
\begin{equation} \label{eq:realtime_calculate_b}
    b = \frac{U_3(s_1) - U_3(s_0)}{f(s_1) - f(s_0)}
\end{equation}
\noindent
where $ s_0 $ and $ s_1 > s_0 $ are selected values of $ s $ for which the PES has already transitioned into the MATM behaviour (we use $ (s_0, s_1) = (3.55, 3.85) $). However, for $ \varphi < \varphi_{\rm cf} = 10^{-5} $, Eqs.~(\ref{eq:realtime_calculate_a}) and (\ref{eq:realtime_calculate_b}) become numerically unstable, and the extrapolation is instead performed using the built-in long-range behaviour of the RKHS method.

When $ R > 3.2 $ \AA, the coefficients $ a(R, \varphi) $ and $ b(R, \varphi) $ both converge to about unity, with numerical noise. We remove this noise from $ a(R, \varphi) $ by transitioning from the numerically calculated coefficients of Eq.~(\ref{eq:a_and_b}) to identically unity using
\begin{equation} \label{eq:transition_a}
    a(R, \varphi) := \overline{\omega}(R; R_{\rm m}, \delta R) a(R, \varphi) + \omega(R; R_{\rm m}, \delta R)
\end{equation}
\noindent
where $ \omega $ and $ \overline{\omega} $ are given by Eqs.~(\ref{eq:cosine_weight_function}) and (\ref{eq:opposite_cosine_weight_function}), with $ R_{\rm m} = 3.2 $ \AA \, and $ \delta R = 0.1 $ \AA. The same filter with the same parameters is applied to the coefficient $ b(R, \varphi) $.
\begin{figure}[h]
	\includegraphics[width=1.0\linewidth]{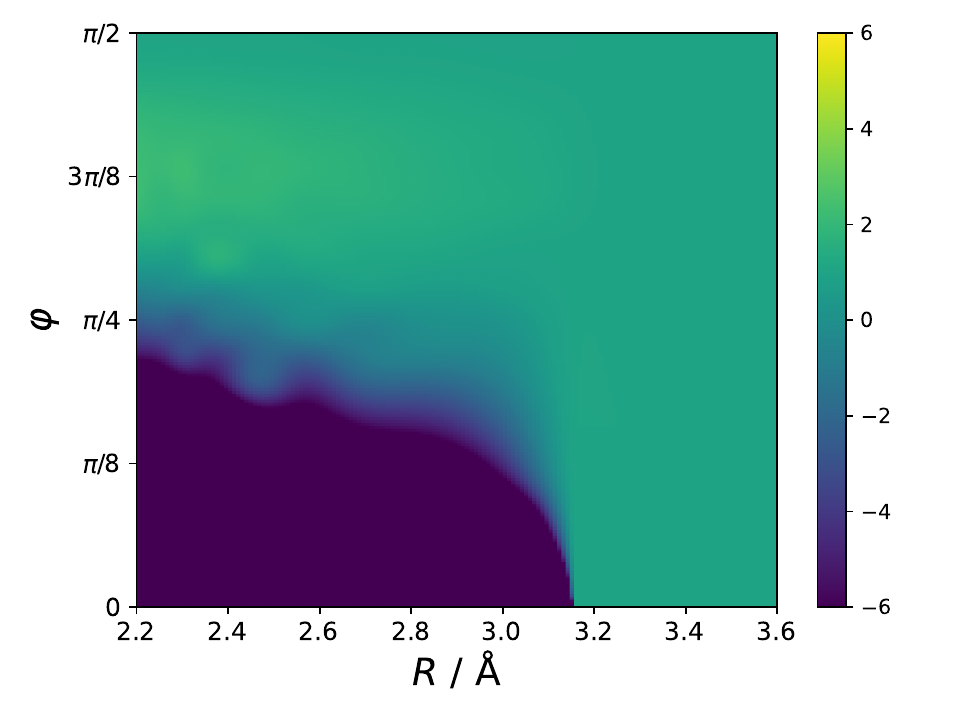}\\
	\includegraphics[width=1.0\linewidth]{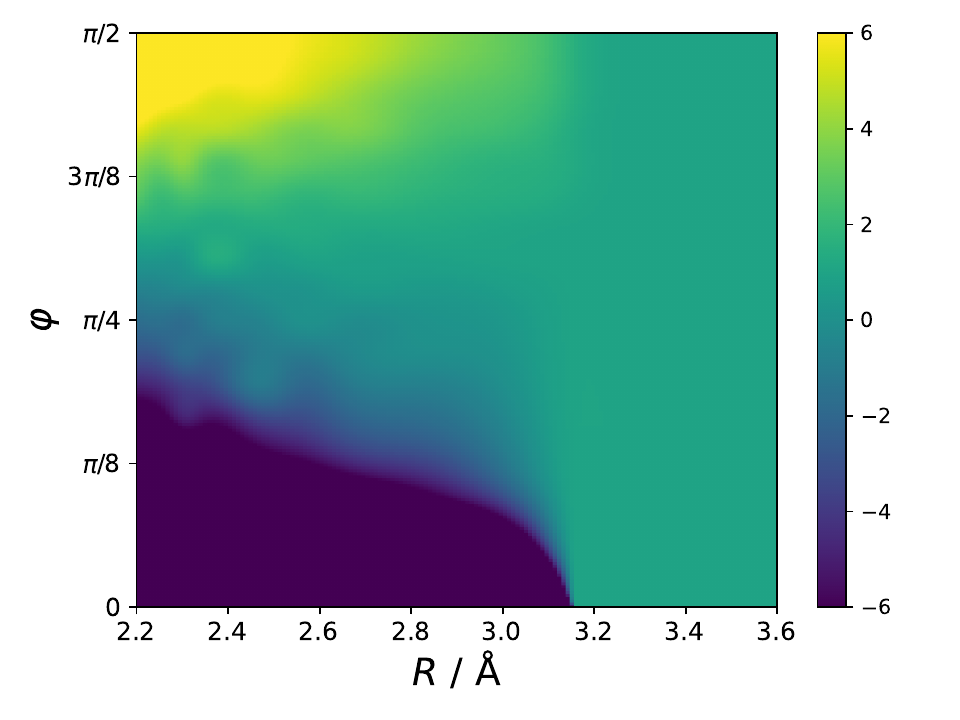}
    \caption{Colormap of $ a(R, \varphi) $ (top panel) and $ b(R, \varphi) $ (bottom panel)   for the final constructed PES, calculated using Eq.~(\protect\ref{eq:a_and_b}).}
	\label{fig4}
\end{figure}

We show colormaps of $ a(R, \varphi) $ and $ b(R, \varphi) $ respectively in Fig.~(\ref{fig4}). For most of the region $ \varphi > 0.9 $, $ a $ is somewhat larger than unity, and peaks with $ a = 2.28 $ near $ \varphi \approx 3 \pi/8 $. For $ R < 3.2 $ \AA, as $ \varphi $ shrinks, the coefficient $ a $ blows up. As expected, $ a $ and $ b $ both become unity past $ R = 3.2 $ \AA. The colormap of $ b(R, \varphi) $ behaves similarly to its counterpart, except when $ R < 3.0 $ \AA \, and $ \varphi > 1.0 $. This region of coordinate space corresponds to small equilateral and ``almost-equilateral'' triangles. The coefficient $ b $ grows fairly large in this region, increasing to a maximum of about $ 8.64 $ when $ (R, \varphi) = (2.2 \mathrm{\ \AA}, \pi/2) $.


\subsubsection{Small R Extrapolation} \label{sec:construct:makepes:small_r}

For most values of $ s $ and $ \varphi $, we can extrapolate the \pHthree interaction energy \red{below} $ R = 2.2 $ \AA \, using an exponential fit
\begin{equation} \label{eq:exponential_fit}
    V_{\rm EX}(R; s, \varphi) = V_3(R_{\rm min}) \exp \left( -c (R - R_{\rm min}) \right) ,
\end{equation}
\noindent
where $ c = c(s, \varphi) $ and (omitting $ s $ and $ \varphi $ on the RHS)
 \begin{equation} \label{eq:exponential_slope}
     c(s, \varphi) = \frac{1}{\Delta R} \ln \left| \frac{ V_3(R_{\rm min} + \Delta R) }{ V_3(R_{\rm min}) } \right|
 \end{equation}
\noindent
with $ R_{\rm min} = 2.2 $ \AA \ and $ \Delta R = 0.05 $ \AA.
However, there are regions in the $ (s, \varphi) $ plane for which $ c $ is very large, and the extrapolation blows up very quickly below $ R_{\rm min} $. These are regions where, in Eq.~(\ref{eq:exponential_slope}), the magnitude of the numerator is a factor of about $ 2 $ or greater than that of the denominator. This problem is often worsened when $ c(s, \varphi) $ is defined using the exact derivative of $ V_3(R) $ instead of a finite approximation. On closer inspection, this behaviour only happens when $ \left| V_3(R_{\rm min}; s, \varphi) \right| $ is already very close to zero (typically $ < 0.05 $ \wvn), and so a small absolute change in energy results in a large relative change in energy. Thus, pre-empting instances where the exponential fit will blow up due to a very large $ c(s, \varphi) $, we transition from the exponential fit to a less accurate but numerically stable linear fit, using
\begin{equation} \label{eq:weighted_exponential_linear}
    V_3(R) = \overline{\omega}(c; c_{\rm m}, \delta c) V_{\rm LI}(R) + \omega(c; c_{\rm m}, \delta c) V_{\rm EX}(R)
\end{equation}
\noindent
where $ c_{\rm m} = 7 $, $ \delta c = 2 $, $ \omega $ and $ \overline{\omega} $ are given by Eqs.~(\ref{eq:cosine_weight_function}) and (\ref{eq:opposite_cosine_weight_function}), respectively, and $ V_{\rm LI}(R) $ is the linear fit below $ R = R_{\rm min} $ whose slope is determined by $ V_3(R_{\rm min}) $ and $ V_3(R_{\rm min} + \Delta R) $. Because the \pHthree interaction energies in regions where the linear fit is used are already very small, as long as the \pHthree interaction potential is not extrapolated too far below $ R_{\rm min} $, the absolute error introduced by using the linear fit will also be very small. When used alongside a (\para)$_2$ pair interaction potential, the error would be negligible. It should be emphasized that this short-range extrapolation along the $ R $ coordinate is not meant to give accurate \pHthree interaction energies below $ R_{\rm min} $. Rather its primary purpose is to make it possible to sample the PES at short distances without it blowing up. Such short range configurations are very unfavourable, and are expected to be rare \red{except} at very high pressures and densities. To ensure a smooth transition of the PES across $ R = R_{\rm min} $, four phantom values of $ R $ with a spacing of $ \Delta R = 0.05 $ \AA $\,$ are created down to $ R = 2.0 $ \AA, and the RKHS method is reapplied.

In the case where both $ R < 2.2 $ \AA \, and $ s > 3.85 $, we first perform the MATM extrapolation along the $ s $-coordinate described in Sec.~\ref{sec:construct:makepes:large_s} to calculate $ V_3(2.2 \, \AA, s, \varphi) $ and $ V_3(2.25 \, \AA, s, \varphi) $. We when use these two energies to extrapolate to values below $ R = 2.2 $ \AA \, with Eqs.~(\ref{eq:exponential_fit}), (\ref{eq:exponential_slope}), and (\ref{eq:weighted_exponential_linear}). Again, we should stress that this extrapolation purposely sacrifices accuracy for stability, and it should not be expected to give highly accurate \pHthree interaction energies.

\subsubsection{Summary} \label{sec:construct:makepes:summary}

\begin{figure}[h]
	\includegraphics[width=1.0\linewidth]{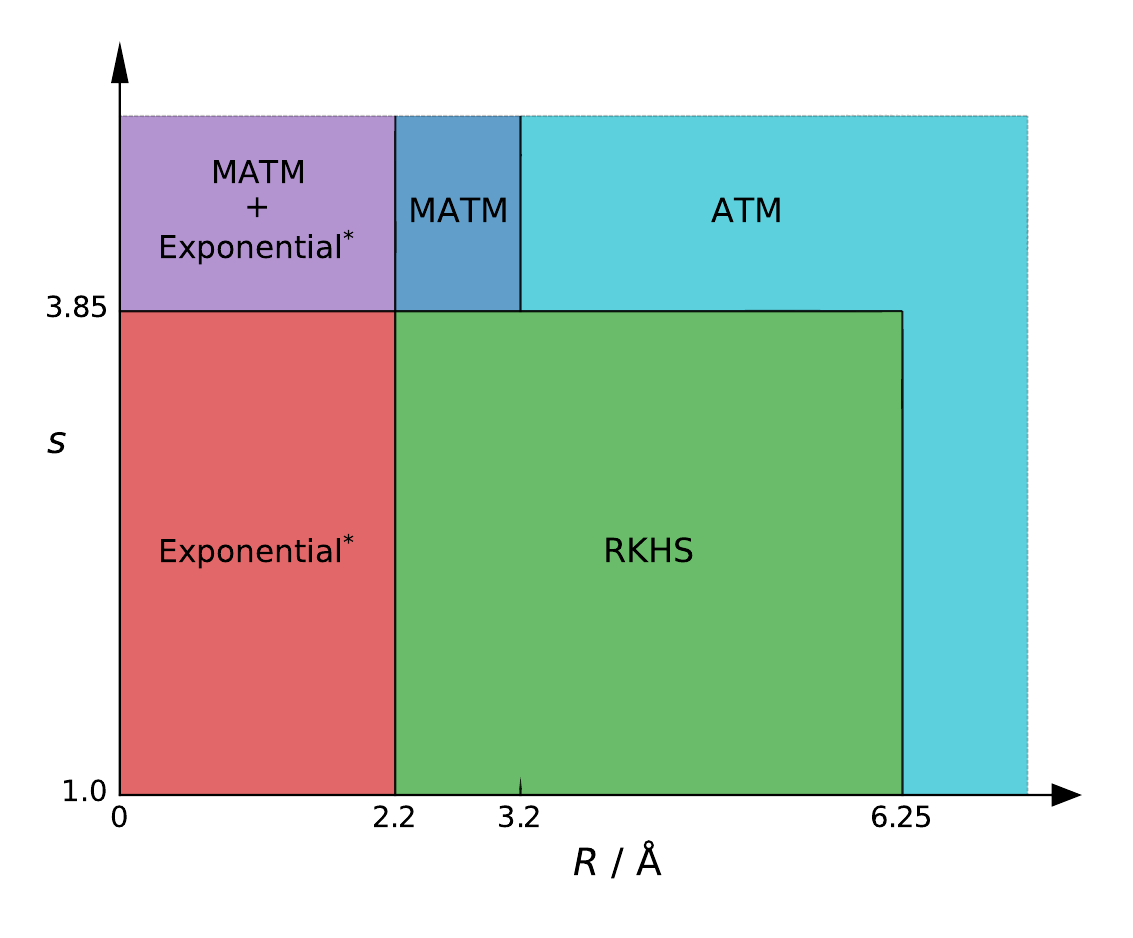}
    \caption{The regions in the $ (R, s) $-plane, labeled with the method used to calculate \pHthree interaction energies within them. Details describing each region are given in the text.}
	\label{fig:paper1_fig5}
\end{figure}

To summarize the construction of the current PES described in this subsection, we show in Fig.~(\ref{fig:paper1_fig5}) the method used to calculate the energies in each region of the $ (R, s) $-plane. With the exception of the MATM potential fit for the case $ \varphi < 10^{-5} $, the $ \varphi $ coordinate is treated the same way in each region depicted in Fig.~(\ref{fig:paper1_fig5}). In the ``ATM'' region, the current PES is identical to the ATM potential. In the ``MATM'' region, the energies are calculated using Eqs.~(\ref{eq:realtime_calculate_a}) and (\ref{eq:realtime_calculate_b}) in Sec.~\ref{sec:construct:makepes:large_s}. In the ``Exponential$^*$'' and ``MATM + Exponential$^*$'' regions, we use the extrapolation procedures described in Sec.~\ref{sec:construct:makepes:small_r}. The ``RKHS'' region is a combination of purely AVTZ-level \textit{ab initio} \pHthree interaction energies and the adjustments to the ATM and MATM potential forms described in Secs.~\ref{sec:construct:makepes:large_r} and \ref{sec:construct:makepes:large_s}. At the border between the ATM and RKHS regions, the RKHS energies are already converged to the functional form of the ATM potential, and the transition is smooth. The same is true for the transition between the MATM and RKHS regions.

In principle, it does not matter in which order the adjustments in Secs.~\ref{sec:construct:makepes:large_r} and \ref{sec:construct:makepes:large_s} are made. In practice, the two adjustments overlap when both $ R $ and $ s $ are large enough, and performing the MATM adjustments in Sec.~\ref{sec:construct:makepes:large_s} first will automatically perform the required adjustments of Sec.~\ref{sec:construct:makepes:large_r} for many values of $ s $ and $ \varphi $. The extrapolation to energies below $ R = 2.2 $ \AA \, can be done independently of either of the aforementioned two adjustments.


\section{Discussion and Results} \label{sec:discuss}


\subsection{Basis Set Comparison} \label{sec:discuss:basiscompare}

We estimate the error due to the truncation of the one-electron basis set by comparing the \pHthree interaction energies calculated using the AVDZ, AVTZ, and AVQZ atom-centred basis sets at several coordinates. In each case, we supplement the atom-centred basis sets with the $ (3s3p2d) $ bond function located at the centre of mass of the trimer. The energies are listed in \red{Tab.~(\ref{tab:basis_energy_compare})}. In essentially each case, we see a much larger jump in energy when we change from the AVDZ to the AVTZ basis set, than from the AVTZ to the AVQZ basis set. We also see that, except for certain cases where the energies are near the point of changing signs the difference between energies calculated using the AVTZ and AVQZ basis sets are either between $1-3 \%$ of each other, or within $ 0.1 $ \wvn of each other. At larger intermolecular separations, the differences are even smaller. These observations are consistent with the previous findings that the \pHthree interaction energies converge very quickly with respect to the choice of one-electron basis set.\cite{threebody:96wind, threebody:08hind}

\begin{figure}[h]
	\includegraphics[width=1.0\linewidth]{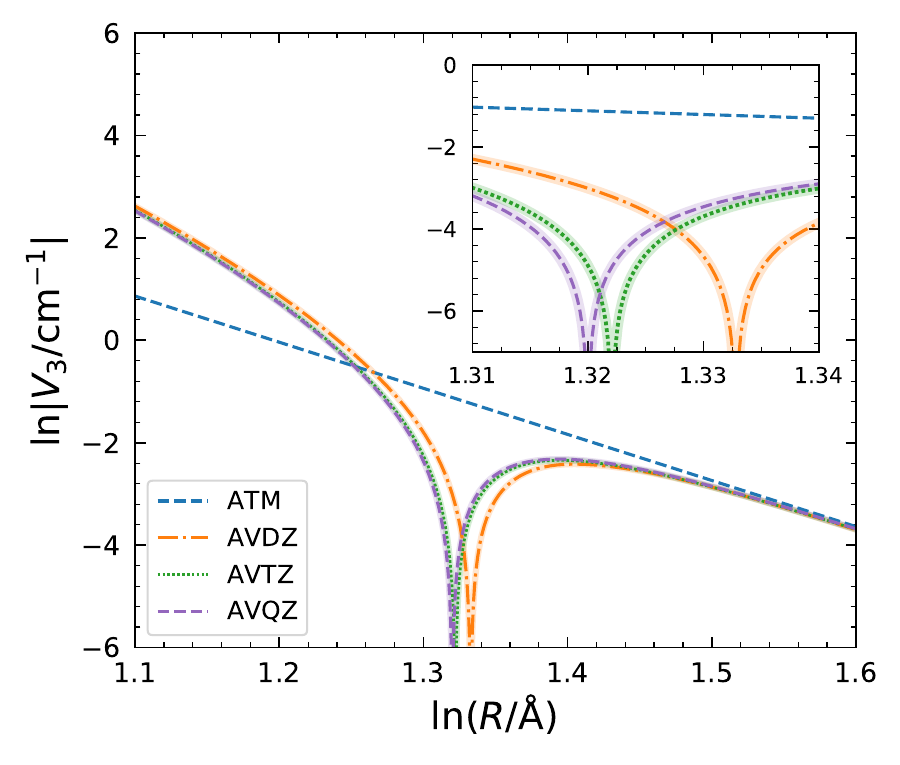}
    \caption{The \pHthree interaction energy for three \para molecules at the corners of an equilateral triangle of side length $ R $. Shown are the ATM potential (straight, blue, dashed), and the RKHS-constructed \textit{ab initio} CCSD(T) energies calculated using the AVDZ (orange, dash-dotted), AVTZ (green, dotted), and AVQZ (purple, dashed) basis sets. The cusp in each \textit{ab initio} curve is located where the curve changes sign from negative to positive. The inset zooms into the region where the cusps are grouped together, and the differences are more pronounced.}
	\label{fig:paper1_fig6}
\end{figure}

In Fig.~(\ref{fig:paper1_fig6}), we compare the interaction energies for three \para molecules arranged in an equilateral triangle configuration, calculated using the AVDZ, AVTZ, and AVQZ atom-centred basis sets (the ``CCSD(T) energy curves''), alongside the ATM potential. The CCSD(T) energy curves were each created using the RKHS method. All three of the CCSD(T) energy curves change from being strongly attractive to \red{weakly} repulsive as the intermolecular distance increases, and the cusp indicates the distance at which the sign of the energy changes. The ATM potential has the expected $ R^{-9} $ decay. The CCSD(T) interaction energies decay exponentially at small $ R $, but trend with and converge to the ATM potential at large $ R $. In the inset, we see that the change in the potential energy curve is much smaller going from using the AVQZ basis set to the AVTZ basis set, than from the AVTZ basis set to the AVDZ basis set. The AVTZ level of treatment should thus be sufficient to describe the \pHthree interaction energy contribution to the PES.

\begin{table} [ht]
    \caption{Interaction energies for the \pHthree trimer (\wvn) at select $ (R, s, \varphi) $ coordinates (with $ R $ in \AA), calculated at the CCSD(T) level using the AVDZ, AVTZ, and AVQZ atom-centred basis sets, supplemented with the $ (3s3p2d) $ bond function at the centre of mass.}
    
    \begin{ruledtabular}
	    \label{tab:basis_energy_compare}
        \begin{tabular}{rrcrrr}
            $R   $ & $s   $ & $\varphi$ &  \mbox{AVDZ}       &  \mbox{AVTZ}       &  \mbox{AVQZ}       \\
            \hline
            $2.20$ & $1.00$ & $0      $ & $  39.3985$ & $  35.1058$ & $  34.4951$ \\
                   &        & $\pi/6  $ & $  21.3524$ & $  18.6636$ & $  18.1592$ \\
                   &        & $\pi/3  $ & $ -94.2664$ & $ -90.8722$ & $ -90.2855$ \\
                   &        & $\pi/2  $ & $-594.3297$ & $-578.3776$ & $-575.2212$ \\
                   & $1.25$ & $0      $ & $   2.6753$ & $   2.0036$ & $   1.9247$ \\
                   &        & $\pi/6  $ & $   2.8044$ & $   2.2194$ & $   2.1194$ \\
                   &        & $\pi/3  $ & $ -10.7030$ & $ -10.0275$ & $  -9.9676$ \\
                   &        & $\pi/2  $ & $-166.8678$ & $-160.3304$ & $-159.1005$ \\
                   & $1.75$ & $0      $ & $  -0.2085$ & $  -0.2181$ & $  -0.2134$ \\
                   &        & $\pi/6  $ & $  -0.1402$ & $  -0.1441$ & $  -0.1445$ \\
                   &        & $\pi/3  $ & $   0.2278$ & $   0.2379$ & $   0.2343$ \\
                   &        & $\pi/2  $ & $  -7.0983$ & $  -6.2079$ & $  -6.1053$ \\
            $2.95$ & $1.00$ & $0      $ & $   0.1483$ & $   0.0160$ & $   0.0041$ \\
                   &        & $\pi/6  $ & $   0.1973$ & $   0.1050$ & $   0.0896$ \\
                   &        & $\pi/3  $ & $  -0.9241$ & $  -0.7669$ & $  -0.7571$ \\
                   &        & $\pi/2  $ & $ -18.1219$ & $ -16.8327$ & $ -16.6135$ \\
                   & $1.25$ & $0      $ & $  -0.1009$ & $  -0.1071$ & $  -0.1082$ \\
                   &        & $\pi/6  $ & $  -0.0568$ & $  -0.0635$ & $  -0.0656$ \\
                   &        & $\pi/3  $ & $   0.0815$ & $   0.0950$ & $   0.0938$ \\
                   &        & $\pi/2  $ & $  -2.1565$ & $  -1.8514$ & $  -1.8101$ \\
                   & $1.75$ & $0      $ & $  -0.0122$ & $  -0.0126$ & $  -0.0135$ \\
                   &        & $\pi/6  $ & $  -0.0095$ & $  -0.0096$ & $  -0.0103$ \\
                   &        & $\pi/3  $ & $   0.0187$ & $   0.0181$ & $   0.0171$ \\
                   &        & $\pi/2  $ & $   0.1580$ & $   0.1692$ & $   0.1685$ \\
            $4.15$ & $1.00$ & $0      $ & $  -0.0240$ & $  -0.0268$ & $  -0.0255$ \\
                   &        & $\pi/6  $ & $  -0.0144$ & $  -0.0150$ & $  -0.0154$ \\
                   &        & $\pi/3  $ & $   0.0219$ & $   0.0222$ & $   0.0221$ \\
                   &        & $\pi/2  $ & $   0.0867$ & $   0.0903$ & $   0.0917$ \\
        \end{tabular}
    \end{ruledtabular}
\end{table}


\subsection{Lebedev Quadrature Comparison} \label{sec:discuss:lebedev}

\begin{figure}[h]
	\includegraphics[width=1.0\linewidth]{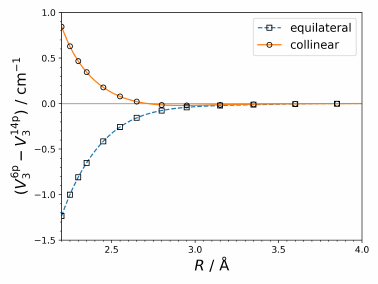}
    \caption{The difference between the $6$-point ($ V_3^{\rm 6p} $) and $14$-point ($ V_3^{\rm 14p} $) Lebedev quadrature averaged \textit{ab initio} \pHthree interaction energies. The energies are calculated using the AVDZ basis set. The curves shown are those for the equilateral triangle configuration of side length $ R $ (orange, solid, circles) and the collinear triangle configuration with spacing $ R $ (blue, dashed, squares).}
	\label{fig:paper1_fig7}
\end{figure}

To assess the error introduced by the use of the $6$-point Lebedev quadrature scheme, we compare \textit{ab initio} \pHthree interaction energies that are spherically averaged using the $6$-point and $14$-point Lebedev quadrature schemes. We do so for two triangle geometries, the $ (R, 1, \pi/2) $ geometry (equilateral triangles of side length $ R $) and the $ (R, 1, 0) $ geometry (three points on a straight line, each a distance $ R $ from its immediate neighbour). The energies are calculated using the AVDZ atom-centred basis, supplemented with the $ (3s3p2d) $ midbond functions. We show in Fig.~(\ref{fig:paper1_fig7}) the difference between the $6$-point and $14$-point integrated energies for these two geometries. Both curves display an exponential decay, although in the case of the collinear configuration this exponential decay is briefly interrupted by a sign change near $ R = 2.7 $ \AA. Below $ R = 2.95 $ \AA, the error from the $6$-point Lebedev quadrature is about $ 0.2 \% $ and $ 2 \% $ for the equilateral and collinear configurations, respectively. For example, at $ R = 2.2 $ \AA, the $6$-point equilateral (collinear) configuration energy is $ -594.329 $ \wvn ($ 39.399 $ \wvn), while the difference between the $6$-point and $14$-point energies is only $ -1.235 $ \wvn ($ 0.845 $ \wvn). Aside from instances near where the interaction energy changes sign, the error decreases even more beyond $ R = 2.95 $ \AA.

For at least the isotropic component of the \pHthree interaction potential, the $6$-point Lebedev scheme is acceptable. Whereas the $6$-point Lebedev quadrature takes the average of only $27$ energies, its $14$-point counterpart requires the energies of $343$ angular orientations. To switch to the $14$-point Lebedev quadrature requires an order of magnitude more electronic structure calculations, while providing relatively little improvement.


\subsection{Effects of Data Mesh Spacing on Potential Energy Surface} \label{sec:discuss:sparsity}

As mentioned previously, the input data mesh used to create the PES is dense at short intermolecular spacings, and gradually increases as the molecules move farther apart. We do this because the AVTZ calculations used in the final constructed PES are fairly expensive. It is both unnecessary and intractable to use the same grid spacing at short intermolecular separations ($ \Delta R = 0.05 $ \AA \, and $ \Delta s = 0.05 $) throughout the entire data mesh. We must find out where, and by what amount, the data mesh can be made sparser while maintaining the accuracy of the interpolation. To do so, we calculated \pHthree interaction energies for a uniform data mesh in the range of $ (R / $\AA$, s, \varphi) \in [2.20, 4.75] \times [1.0, 5.0] \times [0, \pi/2] $, with grid spacings of $ \Delta R = 0.05 $ \AA, $ \Delta s = 0.046 $, and $ \Delta \varphi = \pi/20 $. Overall, the data set was made up of $ 52 \times 100 \times 19 = 98800 $ points. Calculations were performed with the cheaper AVDZ basis set $ + \, \, (3s3p2d) $ midbond function. We construct a PES from this uniform data mesh using the RKHS method.

In a process of trial and error, we remove rows of $ R $ or $ s $ data at different sections of the input mesh, reconstruct the PES with the RKHS method, and compare its accuracy against the PES constructed with the full grid. For $ R < 2.35 $ \AA, we find that the data mesh must remain dense to match the initial accuracy, at $ \Delta R = 0.05 $ \AA. However, as $ R $ increases, we can gradually increase the spacing by $ 0.05 $ \AA \, at different intervals up to $ \Delta R = 0.30 $ \AA \, when $ R > 3.85 $ \AA. A similar trend is seen with the $s$ coordinate.

To compare against the RKHS method, we use the tricubic interpolation method to construct another PES from the dense uniform input data mesh. The implementation of the tricubic interpolation we use is given by Lekien \textit{et al.}\cite{tricubic:05leki} The two methods produce very similar energies (typically less than $ 10^{-6} $ \wvn difference at large $ R $ and $ s $, and less than $ 10^{-2} $ \wvn difference at small $ R $ and $ s $). The only notable exception was between $ R = 2.20 $ \AA \, and $ R = 2.25 $ \AA , where the tricubic PES deviated from the expected exponential behaviour, while the RKHS PES reprodued the expected trend successfully. Ultimately, the RKHS method was chosen over the tricubic interpolation method, because the accuracy of the latter worsened more quickly as the input data mesh spacing increased.

\begin{figure}[h]
	\includegraphics[width=1.0\linewidth]{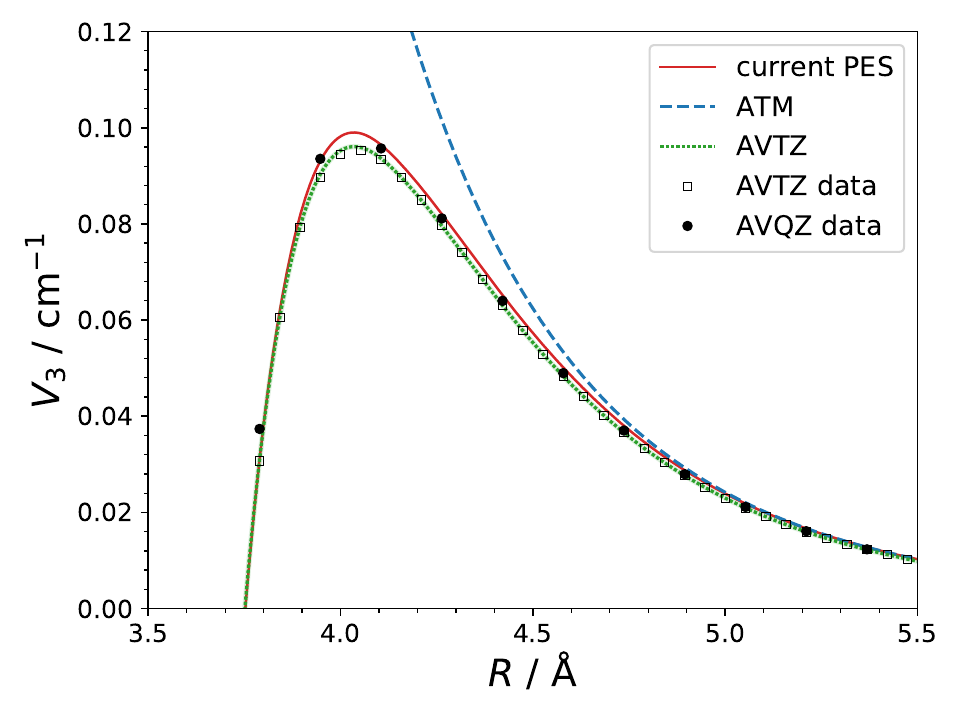}
    \caption{The \pHthree interaction energy for three \para molecules at the corners of an equilateral triangle of side length $ R $. \red{Shown are the ATM potential (blue, dashed), the AVTZ-level RKHS-constructed PES (green, dotted), and the final constructed PES (red, solid). At this distance, the AVTZ-level PES is constructed from training data with a mesh spacing of $ \Delta R = 0.3 $ \AA. Also shown are energies calculated using the AVTZ basis (empty squares, spacing of $ \Delta R = 0.053 $ \AA) and the AVQZ basis (solid circles, spacing of $ \Delta R = 0.159 $ \AA); these energies were not used to construct any shown curve.}}
	\label{fig:paper1_fig8}
\end{figure}

The spacing in the $R$-coordinate for $ R > 3.85 $ \AA \, is fairly large, at $ \Delta R = 0.3 $ \AA, \red{and we want to see how accurately the PES interpolates the (\para)$_3$ interaction energies at this distance.} In Fig.~(\ref{fig:paper1_fig8}), for the equilateral triangle configuration, we plot curves for the current PES, the AVTZ-level RKHS-constructed PES, and the ATM potential. \red{Also shown are AVTZ-level energies with a spacing of $ \Delta R = 0.053 $ \AA \, that act as a ``test set'', unused in the creation of the AVTZ-level PES. Despite being constructed with a mesh spacing about $ 6 $ times greater than the test energies, the AVTZ-level PES provides an excellent interpolation for the independent energy points. Looking at the current PES, we see that as a consequence of correcting the long range behaviour in the AVTZ-level PES, the empirical adjustments also shift the curve closer to the AVQZ energies.} We see similar behaviour for the $s$ coordinate at large $s$, where the spacing is $ \Delta s = 0.3 $. This is expected, as the RKHS method interpolates the $R$ and $s$ coordinates in a similar manner.

\begin{figure}[h]
	\includegraphics[width=1.0\linewidth]{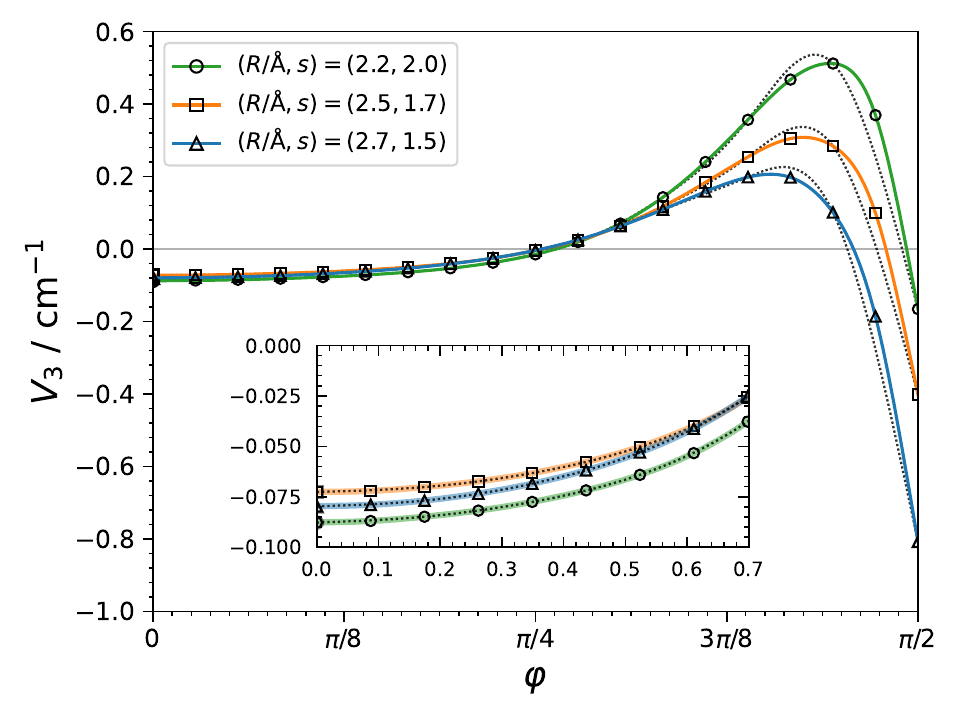}
    \caption{The \pHthree interaction energy curves along the $ \varphi $ coordinate. The RKHS method is applied to the AVTZ-level \textit{ab initio} \pHthree interaction energies, using two different spacings along the $ \varphi $ coordinate. We show the curves for a spacing of $ \Delta \varphi = 5\degree $ [coloured lines, green for $ (R/\mathrm{\AA}, s) = (2.2, 2.0) $, orange for $ (R/\mathrm{\AA}, s) = (2.5, 1.7) $, blue for $ (R/\mathrm{\AA}, s) = (2.7, 1.5) $] and $ \Delta \varphi = 10\degree $ [corresponding black dashed lines]. The markers represent the \textit{ab initio} data at $ \Delta \varphi = 5\degree $ spacing. The inset zooms into the plot at smaller $ \varphi $.}
	\label{fig:paper1_fig9}
\end{figure}

The $\varphi$ coordinate has a finite range, and is interpolated differently by the RKHS toolkit.\cite{rkhs:17unke} To check how the density of points along the $\varphi$ coordinate affects the accuracy of the fit, we reconstruct the current PES, but instead of using $ 19 $ values of $ \varphi $ with a uniform spacing of $ \Delta \varphi = 5\degree $, we remove every other inner point, leaving only $ 10 $ values with a uniform spacing of $ \Delta \varphi = 10\degree $. In Fig.~(\ref{fig:paper1_fig9}), we plot both the original empirically-adjusted PES and the ``sparser'' version of the PES, for three different $ (R, s) $ pairs. For $ \varphi < 3\pi/8 $, both PESs provide similar interpolations, but at larger $ \varphi $, the interpolation becomes more sensitive to the density of the $ \varphi $ coordinate data. The sparser PES performs poorly at predicting the removed energies. We should note that we specifically chose values of $R$ and $s$ for which the relationship between the data spacing and the interaction accuracy is especially pronounced. For most values of $R$ and $s$, such as those where the \pHthree interaction energy is on the order of tens or hundreds of \wvn, the decrease in accuracy from doubling the spacing $ \Delta \varphi $ at large $ \varphi $ is far less drastic, usually on the order of $ < 1 \% $. This is a good sign, as it indicates that the interpolation quality is already well converged with respect to the density of data along the $ \varphi $ coordinate.


\subsection{HCP Lattice Analysis} \label{sec:discuss:hcp}

Consider all the triangles formed by a central ``reference'' \para molecule and any two of its twelve nearest neighbours inside an \hcp lattice, and let $ \alpha $ be the interior angle at the reference \para molecule of this triangle. We find a total of $66$ triangles.
There are, in order of increasing interior angle,
$24$ instances where $ \alpha = 60\degree$ (i.e. equilateral triangles),
$12$ instances where $ \alpha = 90\degree$ (i.e. right-angled triangles),
$3$  instances where $ \alpha = 109.47\degree$,
$18$ instances where $ \alpha = 120\degree$,
$6$  instances where $ \alpha = 146.44\degree$, and
$3$  instances where $ \alpha = 180\degree$ (i.e. three molecules along a straight line).

\begin{figure*}[ht]
	\includegraphics[width=1.00\linewidth]{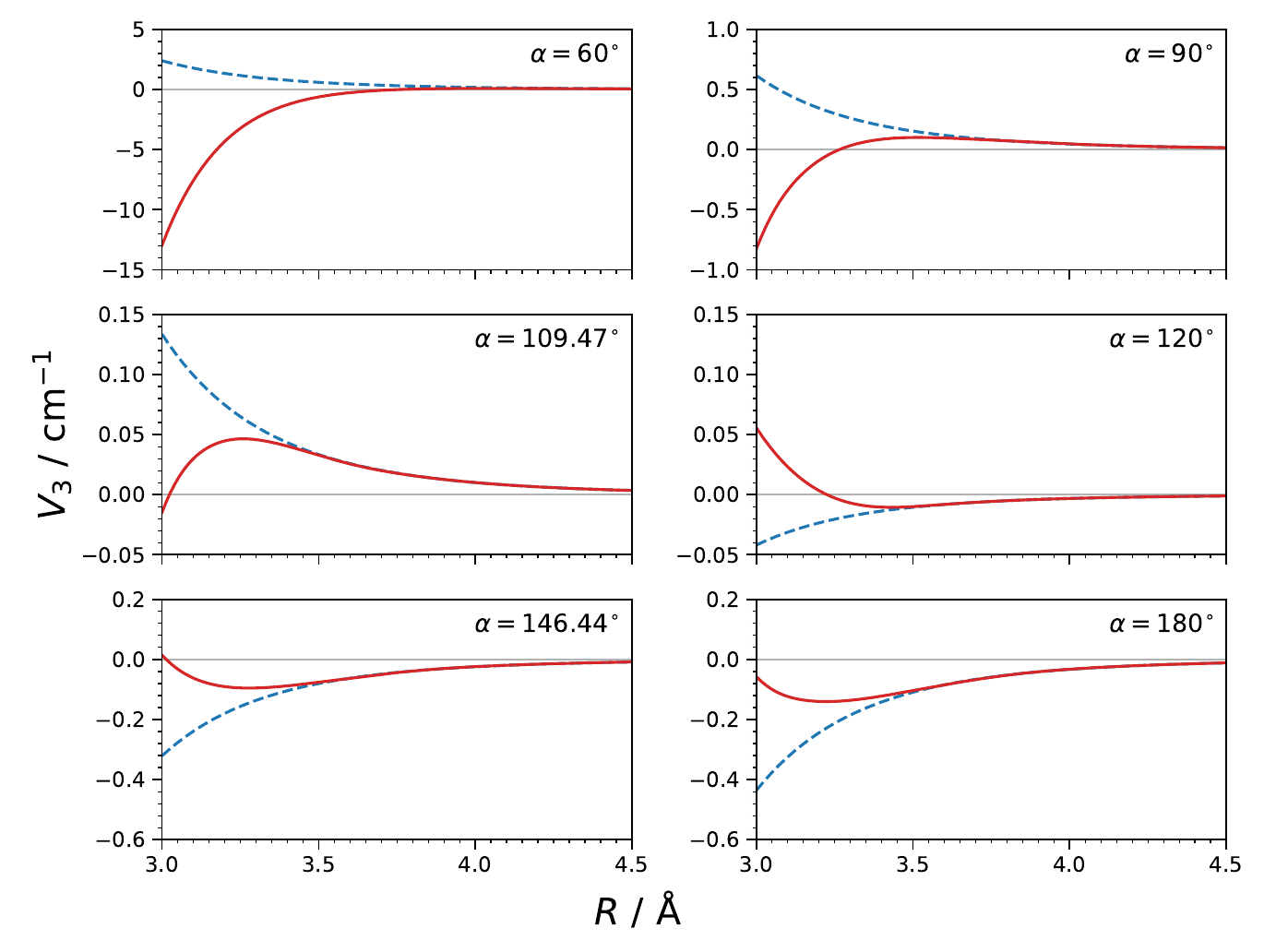}
    \caption{The \pHthree interaction energy using the current PES (solid red line) and the ATM potential (dashed blue line) as a function of the lattice constant $ R $, for the six types of triangles one can make using a central ``reference'' \para molecule and $2$ of its $12$ neighbours in a frozen \hcp lattice. We ask readers to pay attention to the vertical scales of each subplot, as they generally differ for each triangle. Each subplot is labeled using its respective interior angle $ \alpha $ of the triangle at the reference molecule. Note that each subplot corresponds to the \pHthree interaction energy of a single triangle, not the sum of the contributions from all triangles with the same interior angle. For example, the subplot with the label $ \alpha = 60\degree $ is the \pHthree interaction energy for a single equilateral triangle, not all $ 24 $ equilateral triangles of the $ 66 $ triangles considered.}
	\label{fig:paper1_fig10}
\end{figure*}

In Fig.~(\ref{fig:paper1_fig10}), we show the \pHthree interaction energy for each triangle as a function of the lattice constant, for the current PES and the ATM potential. In each instance, the \textit{ab initio} interaction energy diverges from the ATM potential as the lattice constant decreases. Interestingly, for each triplet where the ATM potential becomes more repulsive as the lattice spacing decreases, the \textit{ab initio} \pHthree interaction energy switches sign and becomes progressively more attractive, and vice-versa. This growing deviation shows the inadequacy of the ATM potential to study three-body interactions in \spara at high densities.

At $ R = 3$ \AA, the strength of the \pHthree interaction energy from the equilateral triangle configuration is the greatest of the nearest-neighbour triangles -- it is an order of magnitude greater than that of the right-angled triangle, and at least two orders of magnitude greater than those of the other triangles. Together with its plurality among the aforementioned $66$ triangles, we should expect the equilateral triangle to contribute the majority of the \pHthree interaction energy in an \hcp lattice.

\begin{figure}[h]
	\includegraphics[width=1.0\linewidth]{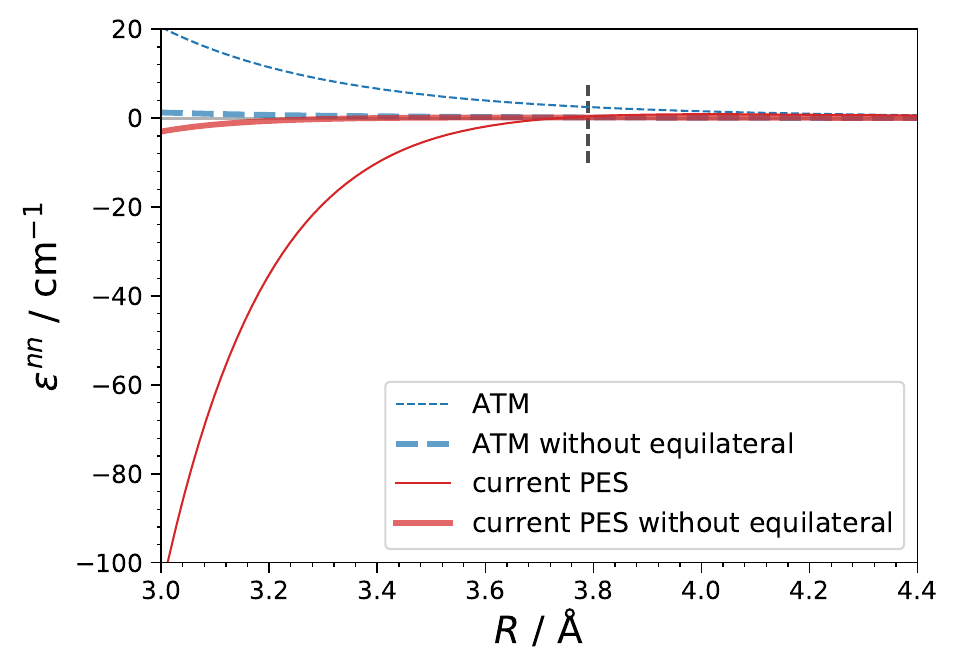}
    \caption{The average \pHthree interaction energy per particle as a function of lattice constant $ R $, between a central reference \para molecule and $2$ of its $12$ nearest neighbours in a frozen \hcp lattice. The curves correspond to the contributions from all $66$ triangles for the current PES (thin solid red line) and the ATM potential (thin dashed blue line), and the contributions from the $42$ non-equilateral triangles for the current PES (thick solid red line) and the ATM potential (thick dashed blue line).}
	\label{fig:paper1_fig11}
\end{figure}

To see if this is the case, in Fig.~(\ref{fig:paper1_fig11}), we compare the contribution to the average interaction energy per particle from all $66$ triangles ($ \epsilon^{nn} $), and from only the $42$ non-equilateral triangles ($ \epsilon^{nn}_{\rm neq} $), using the current PES and ATM potential. These calculations assume a frozen lattice. Select example energies are shown in \red{Tab.~(\ref{tab:hcp_contrib})}. At the ZTZP lattice constant ($ R = 3.79 $ \AA), the difference between $ \epsilon^{nn} $ and $ \epsilon^{nn}_{\rm neq} $ is relatively small ($ 0.40 $ \wvn and $ 0.15 $ \wvn, respectively). However, as the lattice constant decreases, the gap widens considerably, and the equilateral triangles dominate the contribution to the \pHthree interaction energy per particle. This behaviour is caused, not only by the large outlying interaction energy of the equilateral triangle, but also by a partial cancellation of the interaction energies of the other $42$ triangles. In contrast with the current \textit{ab initio} PES, the ATM potential increases the energy per particle.

We can come up with a simple physically intuitive reason for why the interaction energy is the strongest for the equilateral triangle, and weakens as we increase the internal angle $ \alpha $ to form other triangles. In an isotropic \para--\para pair potential, the interaction energy strengthens as the intermolecular spacing decreases. Analogously, the isotropic \pHthree interaction energy strengthens as the total perimeter of the triangle decreases. This trend is apparent when looking at the ATM potential, given by Eq.~(\ref{eq:atm_original}). If we fix the pair distances $ R_{12} = R_{23} = R $, and increase the internal angle $ \alpha_2 $ from $ 60\degree $ to $ 180\degree $, then the distance $ R_{13} $ increases from $ R $ to $ 2R $ and the denominator becomes $ 8 $ times larger. Similarly, we can see from Eq.~(\ref{eq:atm_potential_rsvarphi}) that if we fix $ s $ and $ \varphi $ (and thus constrain ourselves to a set of similar triangles), then the interaction potential increases as the perimeter (which is proportional to $ R $) decreases.

\begin{table}
    \caption{The average \pHthree interaction energy (\wvn) per particle between a central reference \para molecule and $2$ of its $12$ nearest neighbours in a frozen \hcp lattice, for different lattice constants $ R $ (\AA). Shown are the contributions from all $66$ triangles ($ \epsilon^{nn} $) and the $42$ non-equilateral triangles ($ \epsilon^{nn}_{\rm neq} $), for the current PES and the ATM potential.}

    \begin{ruledtabular}
	    \label{tab:hcp_contrib}
        \begin{tabular}{crrrr}
                     & \multicolumn{2}{c}{current PES}                 & \multicolumn{2}{c}{ATM}                         \\
            $ R    $ & $ \epsilon^{nn} $ & $ \epsilon^{nn}_{\rm neq} $ & $ \epsilon^{nn} $ & $ \epsilon^{nn}_{\rm neq} $ \\
            \hline
            $ 2.20 $ & $ -5055.32      $ & $ -428.30                 $ & $ 333.57        $ & $ 22.72                   $ \\
            $ 2.40 $ & $ -2051.03      $ & $ -137.21                 $ & $ 152.44        $ & $  9.47                   $ \\
            $ 2.60 $ & $  -801.71      $ & $  -41.96                 $ & $  74.17        $ & $  4.61                   $ \\
            $ 2.80 $ & $  -300.24      $ & $  -11.93                 $ & $  38.07        $ & $  2.36                   $ \\
            $ 3.00 $ & $  -106.64      $ & $   -3.01                 $ & $  20.46        $ & $  1.27                   $ \\
            $ 3.79 $ & $     0.40      $ & $    0.15                 $ & $   2.50        $ & $  0.16                   $ \\
        \end{tabular}
    \end{ruledtabular}

\end{table}

\subsection{Comparison Against Silvera-Goldman Potential} \label{sec:discuss:silveragoldman}

\begin{figure}[h]
	\includegraphics[width=1.0\linewidth]{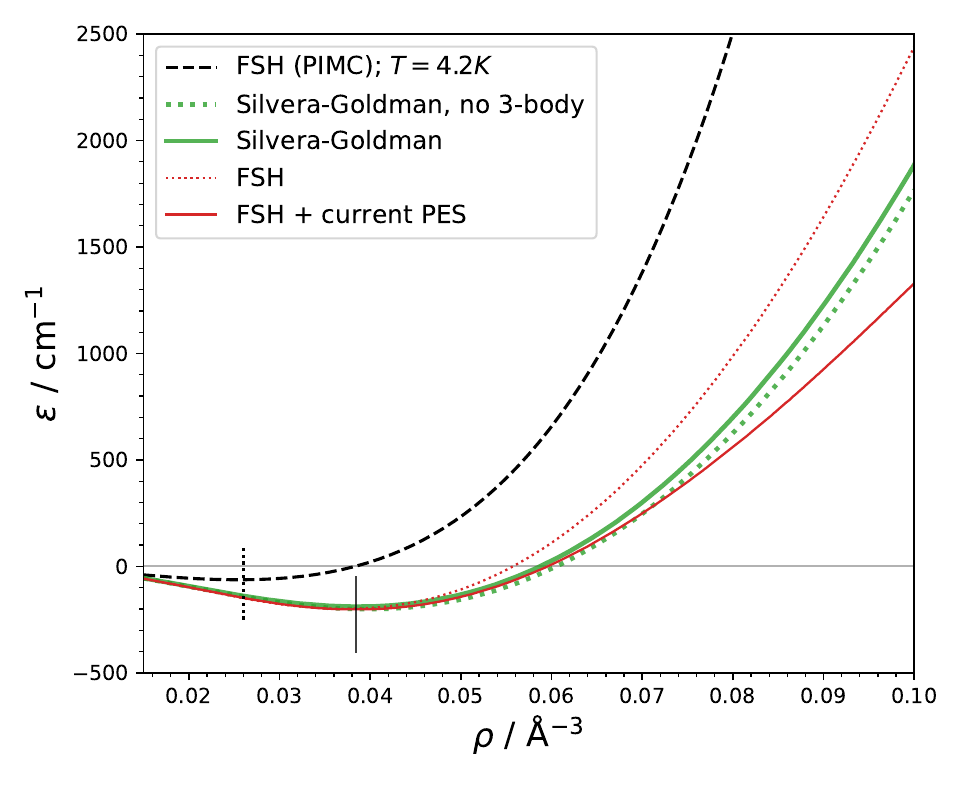}
    \caption{The energy per particle $ \epsilon $ as a function of density $ \rho $ for a frozen \spara \hcp lattice. The curves correspond to the interaction combinations of the FSH potential (red, dotted, thin), the (FSH + current PES) combination (red, solid, thin), the SG potential (green, solid, thick), and the SG potential without its three-body approximation term (green, dotted, thick). Also shown is the curve for the FSH potential calculated from a PIMC simulation (black, dashed) at $ T = 4.2 \, K $.\cite{pathinteg:19ibra} The shift in the energy if the simulation had been done at $ T = 0 \, K $ would not be visible on the scale of this figure. The dotted and solid vertical lines indicate the equilibrium density of the PIMC-simulated FSH curve ($ \rho = 0.0255 $~\AA$^{-3}$), the classical FSH curve ($ \rho = 0.0370 $~\AA$^{-3}$) and (FSH + current PES) curve ($ \rho = 0.0385 $~\AA$^{-3}$).}
	\label{fig:paper1_fig12}
\end{figure}

\begin{figure}[h]
    \includegraphics[width=1.0\linewidth]{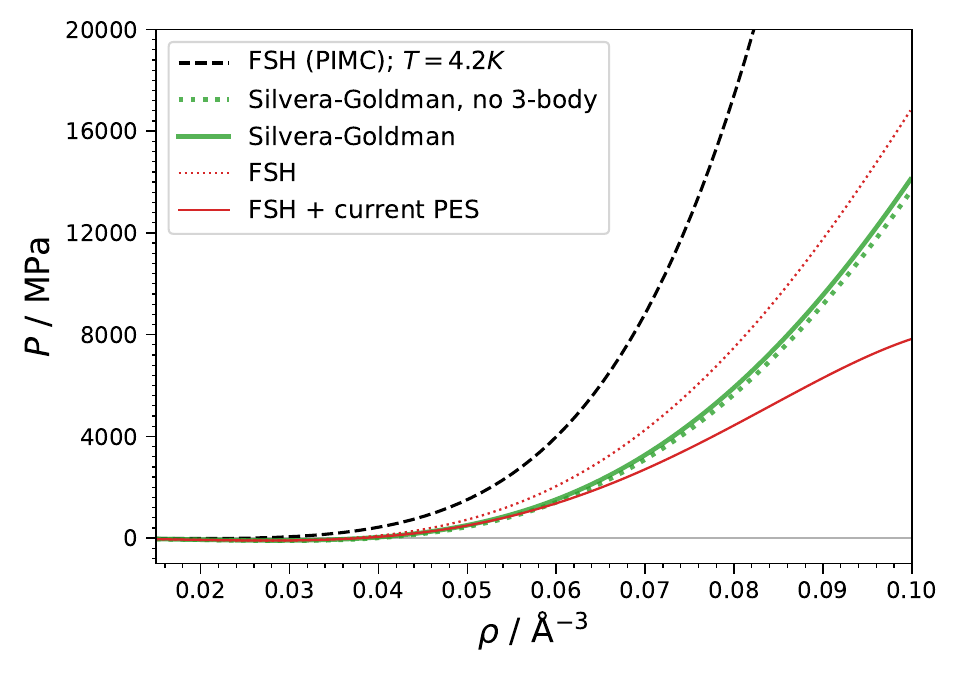}
    \caption{The pressure $ P $ as a function of density $ \rho $ for a frozen \spara \hcp lattice. The curves follow the same labelling in Fig.~(\protect\ref{fig:paper1_fig12}).}
    \label{fig:paper1_fig13}
\end{figure}

How will the introduction of an \textit{ab initio} \pHthree PES influence simulations of condensed \para systems? To estimate its influence, we calculate the energy per molecule $ \epsilon $ as a function of the density $ \rho $ of a \spara \hcp lattice at $ T = 0 $ K with no zero-point motion. From this equation of state (EOS), we calculate the pressure using
\begin{equation}
    P = \eval{\rho^2 \pdv{\epsilon}{\rho}}_T .
\end{equation} We perform this frozen lattice calculation using four different combinations of two-body and three-body \para intermolecular potentials.
These are
the \para--\para FSH potential on its own,
the FSH potential with the current \pHthree interaction PES,
the SG potential, and
the SG potential without its three-body approximation term.
The EOS and pressure curves are shown in Figs.~(\ref{fig:paper1_fig12}) and (\ref{fig:paper1_fig13}), respectively.

Each of the frozen lattice EOS curves predict a much higher ZTZP equilibrium density and a much lower energy per particle than seen experimentally ($ \rho = 0.026 $ \AA$^{-3}$ and $ \epsilon = -62.5 $ \wvn, respectively).\cite{ph2solidexp:80silv, ph2solidtheo:06oper} This discrepancy arises because the frozen lattice calculation does not account for the lattice inflation caused by the zero-point motion of the \para molecules. We see in Fig.~(\ref{fig:paper1_fig13}) that at high densities, compared to the SG potential, the FSH potential on its own predicts a much steeper pressure curve, while the (FSH + current PES) combination decreases the pressure. The former observation is in line with previous findings,\cite{pathinteg:19ibra} where, due to its hard core, the FSH potential greatly overestimates the pressure when $ 0.02 \mathrm{\, \AA}^{-3} < \rho < 0.04 \mathrm{\, \AA}^{-3} $. The latter observation is a good sign, as we know that the SG potential itself begins to overestimate the pressure when extended past $ \rho = 0.1 \mathrm{\, \AA}^{-3} $.\cite{h2pes:12mora, h2pes:13omiy}

\section{Conclusion} \label{sec:conclusion}

We have presented a 3D isotropic \textit{ab initio} \pHthree interaction energy PES, calculated at the CCSD(T) level of theory using an AVTZ atom-centred basis set with an additional $ (3s3p2d) $ midbond function. We have presented in detail the procedure used to construct the PES. The RKHS method interpolates between the data mesh. At long intermolecular separations, the \pHthree interaction energy converges to the ATM potential. In situations where two \para molecules are close together and the third is far away, the \pHthree interaction takes the form of a modified version of the ATM potential. At short intermolecular spacings, the \pHthree interaction energy trends exponentially with $ R $.

The $ (R, s, \varphi) $ coordinate system is a convenient and physically-insightful way to represent the \pHthree system, and we recommend its use for other three-body systems of identical particles. The accuracy of the interpolation was found to be more sensitive to the input data mesh spacings at small values of $ s $ and $ R $, and large values of $ \varphi $. Switching from the $6$-point to the $14$-point Lebedev quadrature provides only minor improvements to the energy.

Among a \para molecule and its nearest neighbours in an \hcp lattice, the equilateral triangle configuration contributes the majority of the \pHthree interaction energy. For many applications, such as computer simulations of solids, one might sufficiently approximate the \pHthree interaction energy using only equilateral and ``almost-equilateral'' triangles. Using the $ (R, s, \varphi) $ coordinate system, this corresponds to creating a PES using an input data mesh with a full range of $ R $ values, but only a small range of $ s $ values past $ s = 1 $, and a small range of $ \varphi $ values below $ \varphi = \pi/2 $. This approximation would greatly reduce the coordinate space one would need to cover to create an adequate three-body PES.

The use of the current \pHthree interaction PES alongside the FSH potential decreases its EOS and pressure for a frozen \hcp lattice at high densities to below that of the SG potential. This indicates, albeit only qualitatively, that the (FSH + current PES) combination is a more powerful model than the SG potential for predicting the properties of many-body \para systems. It has yet to be seen whether the improvements actually occur when the zero-point motion of the \para molecules is taken into account, \red{as} in a path integral simulation. Such a project is currently being carried out in our group, for both \spara and clusters of \para molecules. It is also possible that the influence of the current three-body PES on the structural properties of many-body \para systems can affect these systems' vibrational excitation shifts.\cite{ph2cluster:14faru, pathinteg:19ibra}

\section*{Acknowledgements}
We thank Mih\'{a}ly K\'{a}llay for help with the use of MRCC, and Marcel Nooijen for discussions about electronic structure calculations. P.-N.~R. acknowledges the Natural Sciences and Engineering Research Council (NSERC) of Canada (RGPIN-2016-04403), the Ontario Ministry of Research and Innovation (MRI), the Canada Research Chair program (950-231024), and the Canada Foundation for Innovation (CFI) (project No. 35232). A.~I. acknowledges the support of the NSERC of Canada (CGSD3-558762-2021).

\section*{DATA AVAILABILITY}
\noindent
The data that support the findings of this study are available from the corresponding author upon reasonable request.

\appendix

\section{Conversion of Triangle Side Lengths to Scaled Jacobi Coordinates}

Consider a triangle with side lengths $ R_{12} \le R_{23} \le R_{13} $. Trivially, we set $ R = R_{12} $. Next, we calculate the distance from the origin to molecule $ 3 $ using
\begin{equation}
    r = \sqrt{\frac{1}{2} \left[ R_{13}^2 + R_{23}^2 - \frac{1}{2} R_{12}^2 \right]} .
\end{equation}
\noindent
The coordinates $ \varphi $ and $ s $ are then found using
\begin{equation}
    \cos \varphi = \frac{1}{2} \frac{R_{13}^2 - R_{23}^2}{R_{12} r}
\end{equation}
\noindent
and
\begin{equation}
    s = \frac{2r}{R} \left[ \cos\varphi + \sqrt{3 + \cos^2\varphi} \right]^{-1} .
\end{equation}

\section{Transition Function}

In several instances during the construction of the three-body PES, we find it convenient to use the weight functions
\begin{equation} \label{eq:cosine_weight_function}
    \omega(x; x_{\rm m}, \delta x) =
        \begin{cases}
            0                                          & x \le x_0,     \\
            \frac{1}{2} \left[ 1 - \cos(\pi k) \right] & x_0 < x < x_1, \\
            1                                          & x \ge x_1,     \\
        \end{cases}
\end{equation}
\noindent
and
\begin{equation} \label{eq:opposite_cosine_weight_function}
    \overline{\omega}(x; x_{\rm m}, \delta x) = 1 - \omega(x; x_{\rm m}, \delta x)
\end{equation}
\noindent
where we use $ x_0 = x_{\rm m} - \frac{1}{2} \delta x $, $ x_1 = x_{\rm m} + \frac{1}{2} \delta x $, and $ k = (x - x_0)/\delta x $.

\section*{References}

\bibliography{./biblio}

\end{document}